\documentclass[fleqn,usenatbib]{mnras}

\usepackage{graphicx}	
\usepackage{array}
\usepackage{amsmath}	
\usepackage{footnote}

\title[Short]{Mid-Infrared variability of $\gamma$-ray emitting blazars}

\author[Anjum et al.]{Ayesha Anjum,$^{1}$\thanks{E-mail: astro.pray@gmail.com (KTS)}
C. S. Stalin,$^{2}$
Suvendu Rakshit,$^{3}$  
S. B. Gudennavar,$^{1}$ Alok Durgapal$^{4}$
\\ \\
$^{1}$Department of Physics and Electronics, CHRIST (Deemed to be University), Hosur Road, Bengaluru-560029, India, 360029\\
$^{2}$Indian Institute of Astrophysics, Block II, Koramangala, Bengaluru, India, 560034\\
$^{3}$Finnish Centre for Astronomy with ESO (FINCA), University of Turku, Quantum, Vesilinnantie 5, 20014 University of Turku, Finland \\
$^{4}$Center of Advanced Study, Department of Physics, D. S. B. Campus, Kumaun University, Nainital 263002, India
}

\date{Accepted XXX. Received YYY; in original form ZZZ}

\pubyear{2020}

\begin{document}
\label{firstpage}
\pagerange{\pageref{firstpage}--\pageref{lastpage}}
\maketitle

\begin{abstract}
Using data from the Wide-field Infrared Survey Explorer we studied the 
mid infrared 3.4 $\mu$m (W1-band) and 4.6 $\mu$m (W2-band) flux variability 
of $\gamma$-ray emitting blazars. Our sample  
consists of 460 flat spectrum radio quasars (FSRQs) 
and 575 BL Lacertae (BL Lac) objects.  On intra-day 
timescales, the median amplitude of variability ($\sigma_{m}$) for FSRQs is 0.04$^{+0.03}_{-0.02}$ mag and 0.05$^{+0.03}_{-0.02}$ mag in W1 and W2 bands. For BL Lacs we found median $\sigma_{m}$ in W1(W2) bands of 0.04$^{+0.01}_{-0.02}$ (0.04$^{+0.02}_{-0.02}$) mag. On long timescales, for FSRQs we found a median 
 $\sigma_{m}$ of 0.44$^{+0.28}_{-0.27}$ mag and 0.45$^{+0.27}_{-0.27}$ mag in 
W1 and W2 bands, while for BL Lacs the median values are 0.21$^{+0.18}_{-0.12}$ mag and 0.22$^{+0.18}_{-0.11}$ mag in W1 and W2 bands. From statistical tests, we found FSRQs to show larger $\sigma_m$ than BL Lacs on both intra-day and long timescales. Among blazars, low synchrotron peaked (LSP) sources showed larger $\sigma_m$ compared to intermediate synchrotron peaked (ISP) and high synchrotron peaked (HSP) sources.The larger $\sigma_{m}$ seen in FSRQs relative to BL Lacs on both intra-day and long timescales could be due to them having the most powerful relativistic jets and/or their  mid infrared band coinciding with the peak of the electron energy distribution. BL Lacs have low power jets and the observational window too traces the emission from low energy electrons, thereby leading to low $\sigma_{m}$. In both FSRQs and BL Lacs predominantly  a  bluer when brighter behaviour was observed. No correlation is found between $\sigma_m$ and black hole mass.
\end{abstract}




\section{Introduction}
 
The extragalactic $\gamma$-ray sky is dominated by the blazar category of 
active galactic nuclei (AGN) as evident from the {\it Fermi}-$\gamma$-ray space 
telescope \citep{2009ApJ...697.1071A} observations since its launch in 2008 \citep{2019arXiv190210045T}. AGN 
are believed to be powered by accretion of matter onto supermassive black holes 
located at the centres of galaxies \citep{1969Natur.223..690L,1984ARA&A..22..471R}. 
Blazars, a peculiar category of AGN,  which comprises both flat spectrum radio-quasars (FSRQs) and BL Lacs 
objects (BL Lacs) emit radiation over the entire accessible electromagnetic 
spectrum extending from low energy radio to high energy TeV $\gamma$-ray energies 
predominantly by non-thermal emission processes. This classification of blazars 
into FSRQs and BL Lacs is based on the rest-frame equivalent width (EW) of their optical 
emission lines with BL Lacs having EW $<$ 5 \AA  \citep{1991ApJ...374..431S,1991ApJS...76..813S}. 
Presence of either weak emission lines or featureless 
spectrum in BL Lacs is thought to be because of relativistic beaming, owing to 
their relativistic jets being aligned closely to the observer. However, according 
to \citet{2011MNRAS.414.2674G}, blazars can be divided into FSRQs and BL Lacs based 
on the luminosity of their broad emission lines ($L_{BLR}$) relative to the Eddington 
luminosity ($L_{Edd}$) with BL Lacs having ${L_{BLR}}/{L_{Edd}} < 5 \times 10^{-4}$, where $L_{Edd} = 1.3 \times 10^{38} \left(\frac{M_{BH}}{M_{\odot}}\right)$  erg sec$^{-1}$, $M_{BH}$ is the mass of the black hole.  Apart from the noticeable differences in the optical spectra, both FSRQs and 
BL Lacs have flat radio spectra at GHz frequencies with the spectral index 
$\alpha$ $<$ 0.5 ($S_{\nu} \propto \nu^{-\alpha}$) and show superluminal motion in the radio 
band \citep{2005AJ....130.1418J}. They exhibit rapid flux variations over the 
electromagnetic spectrum on a range of timescales from minutes to years 
\citep{2015ApJ...811..143P,2016ApJ...817...61P,2004MNRAS.350..175S,2017MNRAS.466.3309R}. 
They are highly polarized in optical which is also found to vary with time 
\citep{2005A&A...442...97A,2017ApJ...835..275R,2019MNRAS.486.1781R}. However, in the large scale jet 
structure, FSRQs and BL Lacs do differ with FSRQs being the beamed counterparts 
of the luminous Fanaroff-Riley type II (FR II) radio galaxies \citep{1974MNRAS.167P..31F}  
and BL Lacs being the beamed counterparts of the less luminous FR I radio galaxies.

The broad band spectral energy distribution (SED) of blazars has a typical two 
hump structure. The low energy hump peaking between infrared and X-rays is known 
to result from synchrotron emission process. The high energy hump peaks in the 
MeV to TeV range and its origin is a matter of intense debate, and two competing 
models are available in the literature to explain the high energy hump in blazars. 
In the one zone leptonic emission model, the high energy hump is explained by 
inverse Compton (IC) process. The seed photons for the IC scattering can either originate from within the jet, called the synchrotron self 
Compton \citep{1981ApJ...243..700K, 1985ApJ...298..114M,1989ApJ...340..181G} or 
external to the jet, called the external 
Compton process \citep{1987ApJ...322..650B,1989ApJ...340..162M,1992A&A...256L..27D}. 
Alternatively, the high energy hump can also be explained by hadronic process 
\citep{2013ApJ...768...54B}. Based on the peak of the synchrotron emission, 
blazars are further classified \citep{2010ApJ...716...30A} into low synchrotron peaked 
blazars (LSP; $\nu^{S}_{peak} < 10^{14}$ Hz), intermediate synchrotron peaked 
blazars (ISP; $10^{14}$ Hz < $\nu^{S}_{peak} < 10^{15}$ Hz) and high synchrotron 
peaked blazars (HSP; $\nu^{S}_{peak} > 10^{15}$ Hz).

Blazars have been studied extensively for flux variability in different 
wavelengths at different timescales such as the optical \citep{2006MNRAS.366.1337S}, 
X-ray \citep{2017MNRAS.466.3309R}, UV \citep{1991ApJ...372L...9E,1992ApJ...401..516E} 
and radio \citep{2018MNRAS.480.5517L}. However, our knowledge on the infrared 
variability characteristics of blazars is very 
limited \citep{2016ApJ...817..119K}, though few individual sources have been 
studied \citep{2015RAA....15.1784Z,2015MNRAS.450.2677C,2018RNAAS...2..130G}. 
Recently, \cite{2018Ap&SS.363..167M} investigated the long term mid infrared variability of blazars using about 
four years of data, however, there is no report yet in literature on their mid infrared variability
characteristics on intra-day timescales. Furthermore, there is no 
comparative study of the mid infrared variability of the 
different sub-classes of blazars such as LSP, ISP and HSP available in the literature. 
Studies of infrared variability 
are indeed important to understand the contribution of jet, accretion disk and torus to the 
observed infrared emission. As the mid infrared variability study of 
blazars is very limited, it is important to carryout such a study 
for a clear picture of their mid infrared 
variability.  We have therefore carried out a systematic study on the mid infrared 
flux variability of a large sample of blazars with the following objectives (a) 
to characterize the mid infrared variability characteristics of $\gamma$-ray 
emitting blazars in general, on both intra-day timescales and long timescales (b) to see for similarities and differences between 
the mid infrared variability characteristics of FSRQs and BL Lacs and (c) to have a 
comparative analysis of the mid infrared variability characteristics of LSP, ISP and HSP blazars.
Our sample of blazars for this study  was taken from the third catalog of AGN 
by \citet{2015ApJ...810...14A}. This is the first 
systematic study of the mid infrared flux variability characteristics of
$\gamma$-ray emitting blazars on intra-day timescales. We present the sample and data used in this 
study in Section 2 and the analysis in Section 3. The results are discussed in 
Section 4 followed by the  summary in the final 
section.

\section{Sample and Data}

The sample of blazars used in this study was taken from the third catalog of 
AGN detected by the {\it Fermi} Large Area Telescope \citep[3LAC,][]{2015ApJ...810...14A}. 
Our initial sample consists of 1099 sources of which 467 are FSRQs and 632 are 
BL Lacs. As the prime motivation of this work is to characterise the mid infrared 
variability of {\it Fermi} blazars, we searched for mid infrared counterparts to our 
initial sample of blazars in the Wide-field Infrared Survey Explorer 
\citep[WISE;][]{2010AJ....140.1868W} all sky catalog. Since its launch and 
until 2012, WISE mapped the sky in 4 mid infrared bands namely, W1 (3.4 $\mu$m), 
W2 (4.6 $\mu$m), W3 (12 $\mu$m) and W4 (22 $\mu$m). Its cryogenics 
failed in 2012 and post 2012, WISE carried out observations in only 2 bands, 
W1 and W2. The images from WISE observations have spatial resolution of 6.1, 
6.4, 6.5 and 12 arcsec, in W1, W2, W3 and W4 bands respectively. With WISE making 
about 15 orbits/day, it is natural to get many photometric points on a single 
object in a day. The data from WISE was released in two separate catalogs namely, 
the AllWISE\footnote{http://wise2.ipac.caltech.edu/docs/release/allwise/} 
source catalog (Prior to the cryogenic failure) and NEOWISE catalog\footnote{http://wise2.ipac.caltech.edu/docs/release/neowise/}
(Post Cryogenic failure). The magnitudes given in WISE are in the Vega system 
without any corrections for Galactic extinction. Thus, the multi-epoch photometry 
available in AllWISE and NEOWISE catalogs can be used to investigate the mid infrared 
flux variability properties of {\it Fermi} blazars.

\begin{figure}
\includegraphics[scale=0.42]{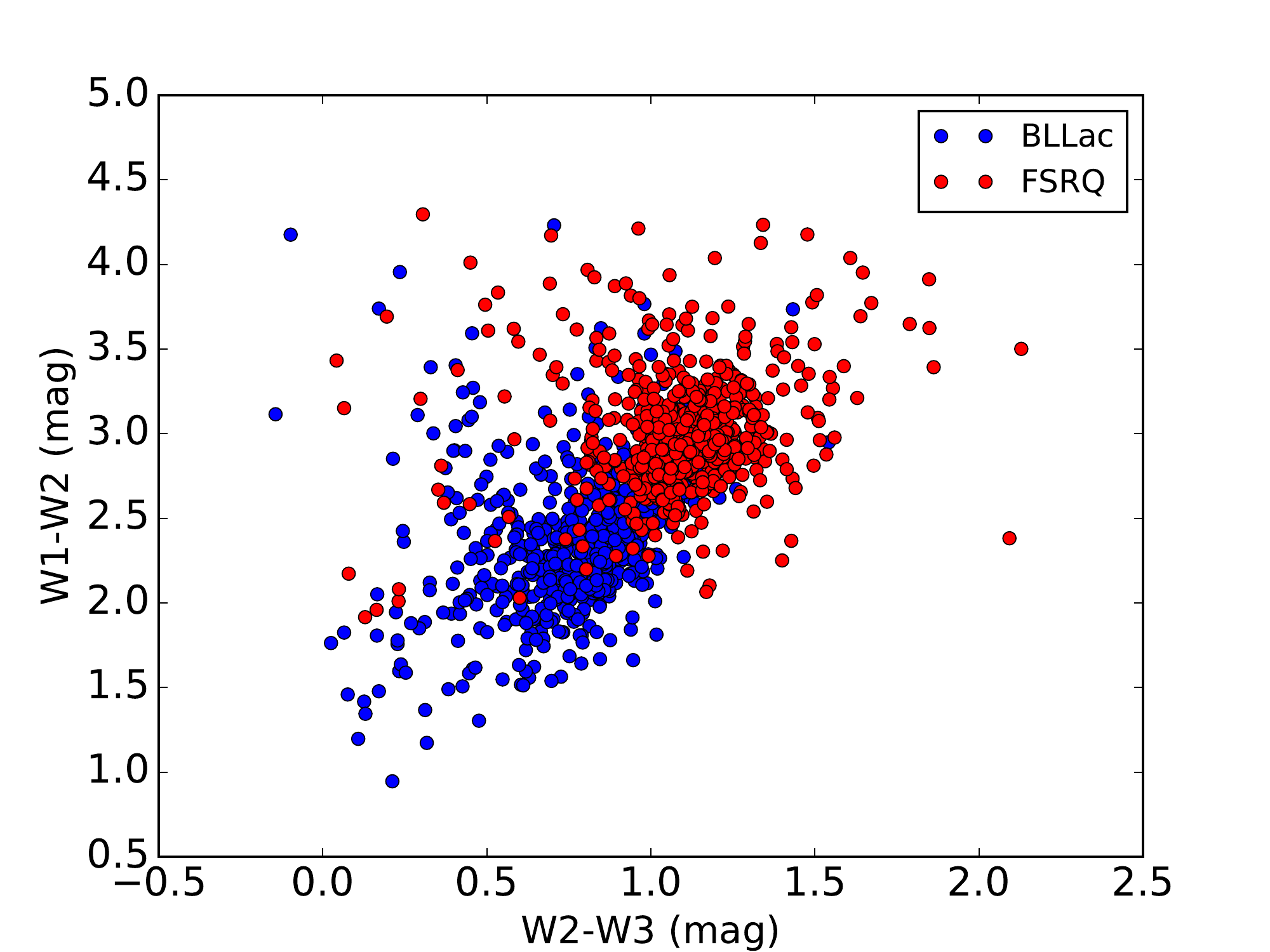}
\caption{The distribution of the sources used in this study in the WISE colour-colour diagram.
FSRQs (red filled circles) and BL Lacs (blue filled circles) are  separated in the 
W1-W2 v/s W2-W3 colour-colour diagram.}
\label{Figure:fig-1}
\end{figure}

\begin{figure}
\hspace*{-0.5cm}\includegraphics[scale=0.42]{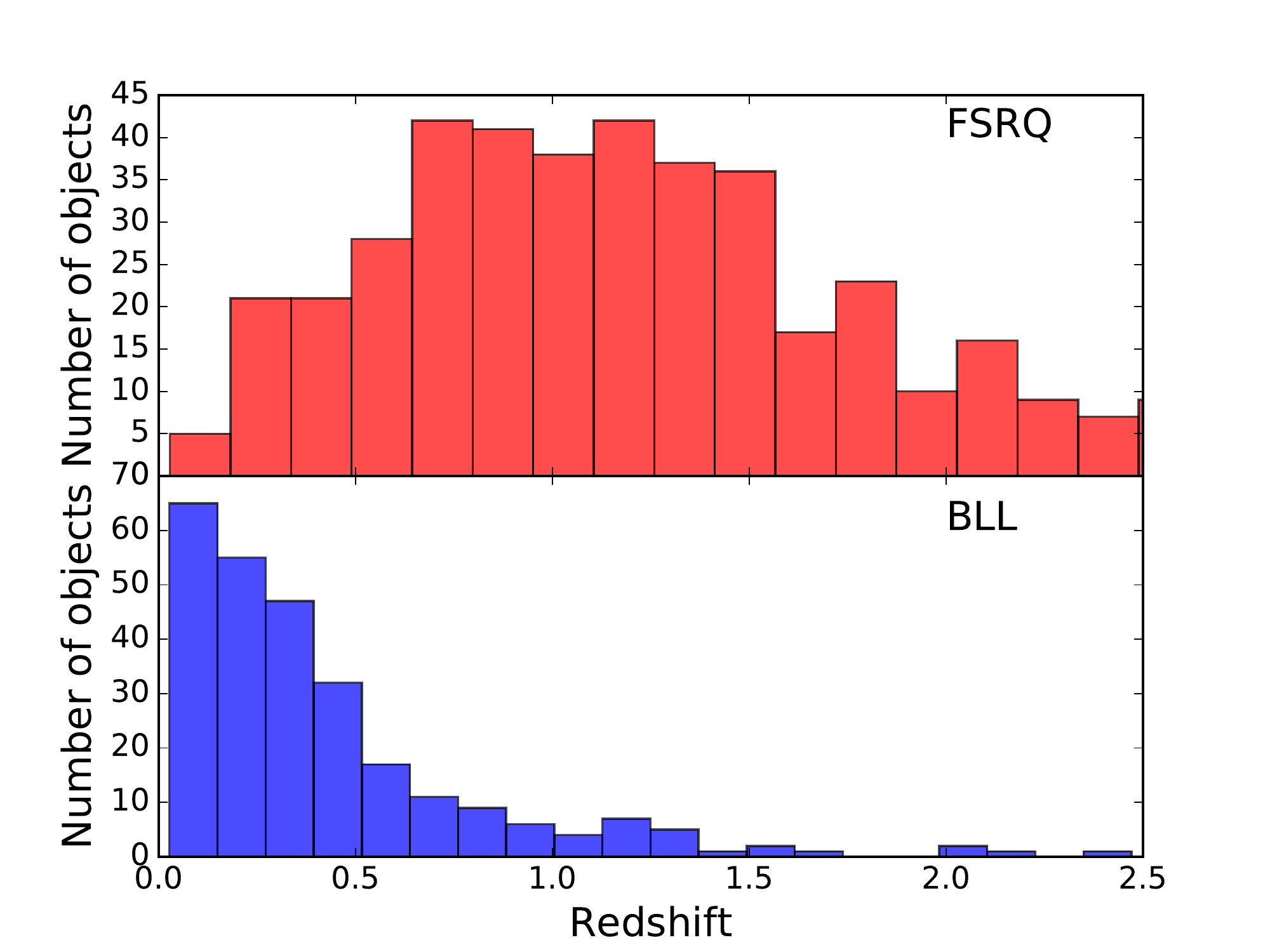}
\caption{The redshift distribution of FSRQs and BL Lacs used in this work.}
\label{Figure:fig-2}
\end{figure}

We cross-correlated our initial sample of 1099 blazars selected
from \citet{2015ApJ...810...14A} with the AllWISE source catalog with a search 
radius of 2 arc second. Our cross correlation yielded 1035 sources. The 
distribution of our sample of 1035 sources in the WISE colour-colour diagram is 
shown in Figure~\ref{Figure:fig-1}. These 1035 sources form our sample for mid infrared 
variability study. Of these 1035 sources, 575 are BL Lacs and 460 are FSRQs. FSRQs 
cover the redshift from 0.189 to 3.10, while BL Lacs cover the redshift between 
0.034 and 1.72. The distribution of redshifts taken from \cite{2015ApJ...810...14A} for our final sample of FSRQs and BL Lacs are given in Figure~\ref{Figure:fig-2}. In our sample, about 10\% of FSRQs and 50\% of BL Lacs do not have redshift measurements. Further dividing our sample, based on the peak of the synchrotron emission in their broad band SED, we have 565 LSPs, 207 ISPs and 243 HSP sources. A total of 20 sources in our sample, do not have sub-classification in \cite{2015ApJ...810...14A}. The summary of the different types of sources used in this study is given in Table~\ref{Table:table-1}. 

\begin{figure}
\hspace*{-0.5cm}\includegraphics[scale=0.25]{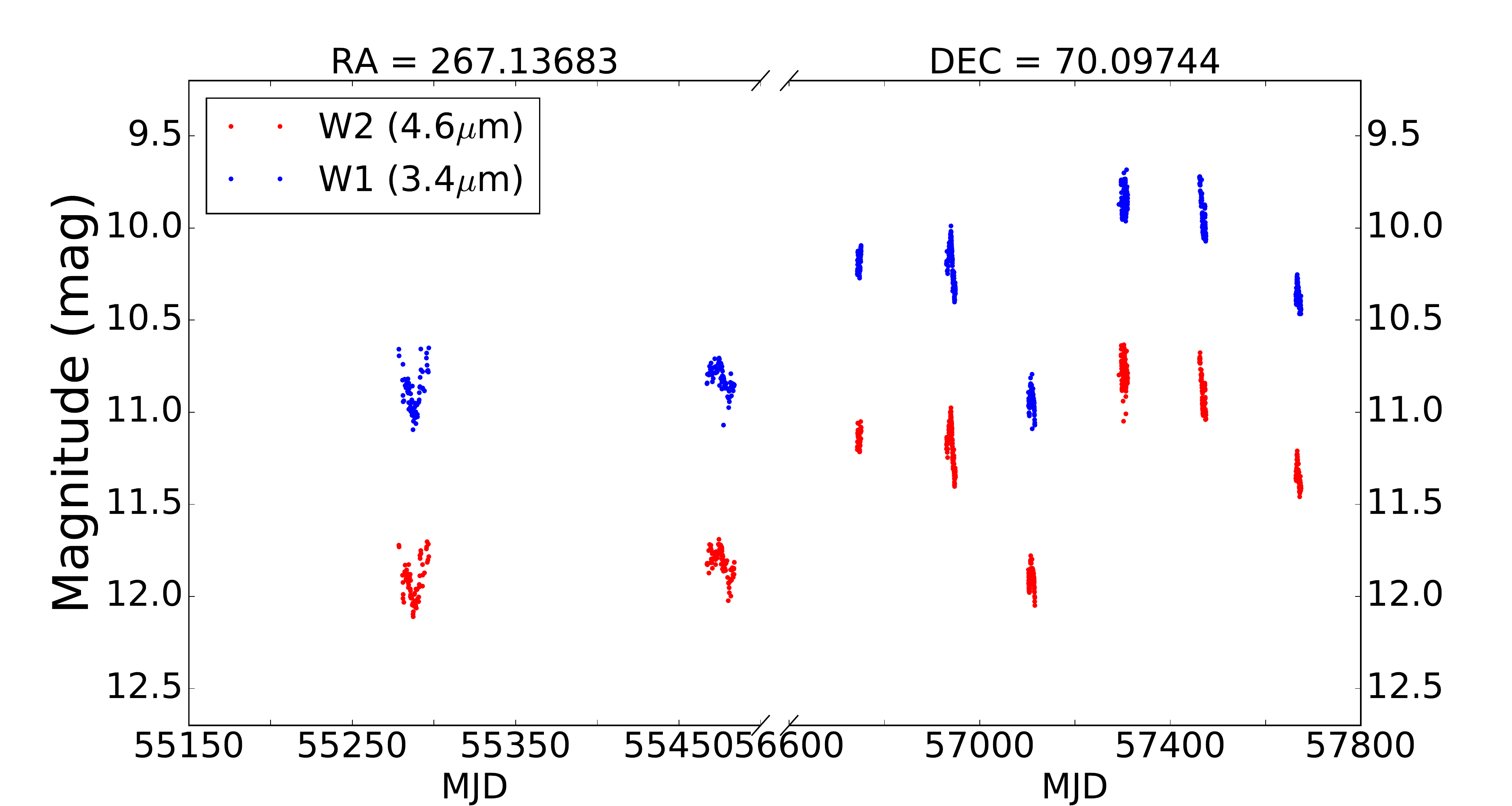}
\caption{Sample light curves spanning about seven years for the BL Lac (J1748.6+7005) in W1 (red) and W2 (blue) bands.}
\label{Figure:fig-3}
\end{figure}

\begin{figure}
\includegraphics[scale=0.36]{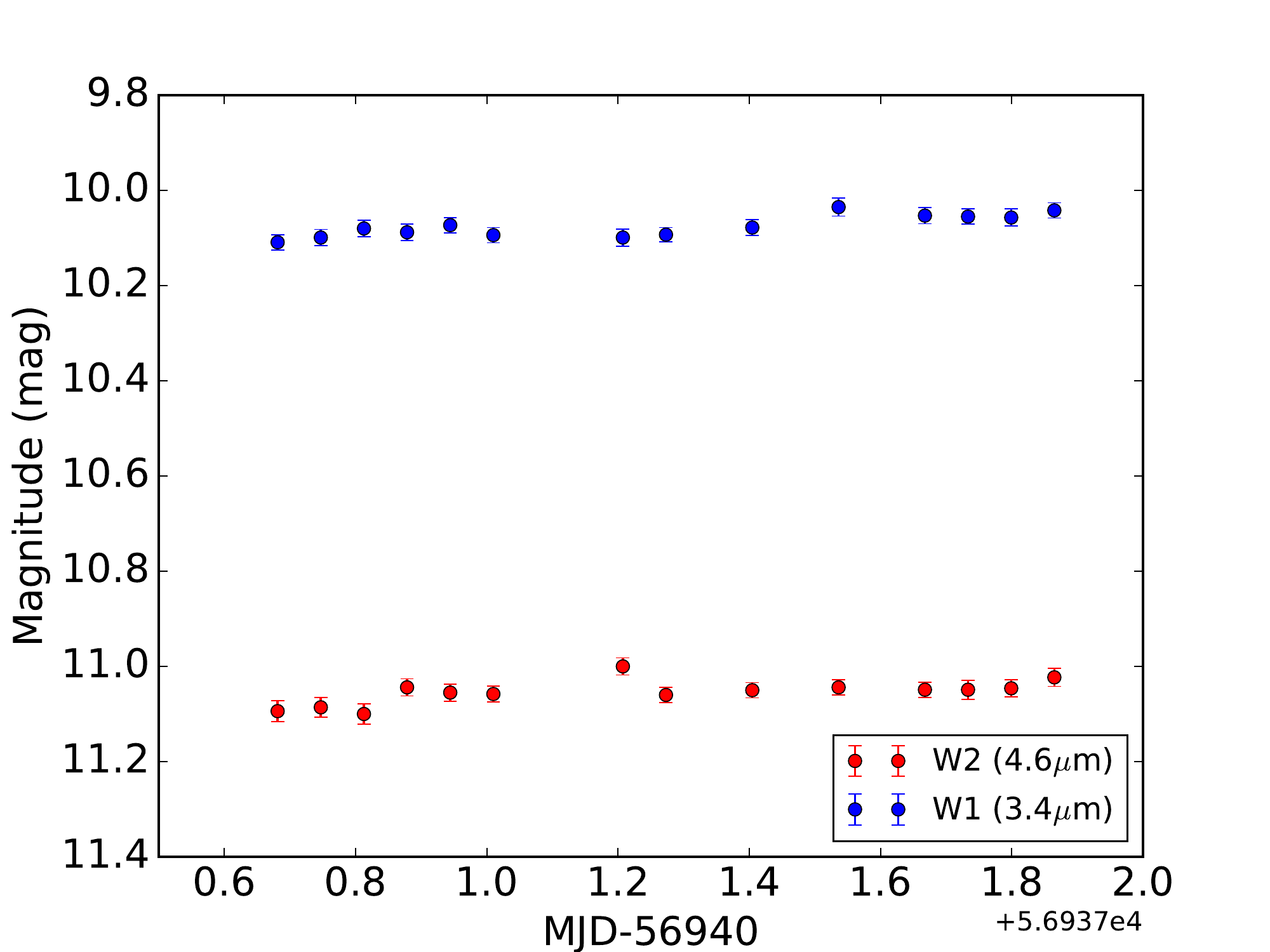}
\hspace{0.8cm}
\caption{Expanded one day light curves in W1 (red) and W2 (blue) bands for the 
source J1748.6+7005 shown in Figure \ref{Figure:fig-3}.}
\label{Figure:fig-4}
\end{figure}

For these 1035 sources, we also looked into the availability of data in
NEOWISE \citep{2011ApJ...731...53M}.  Of these 1035 sources, NEOWISE data was 
available for 914 sources. For sources having data only in AllWISE, the observed 
duration spans a period of about a year between MJD 55203 and MJD 55593. 
However, for sources that have observations both in AllWISE and NEOWISE the 
duration of  observations covers a period of about 7 years between MJD 55203 and 
MJD 57735. Also, for observations within a day, there can be points as large as 
the number of orbits made by WISE, thereby enabling us to study variability on
both intra-day timescales (of the order of hours) and long timescales (of 
the order of years). A sample light curve of the  source (a BL Lac J1748.6+7005 at $z$=0.77)  in W1 and W2 bands 
spanning about 7 years of observation is given in Figure~\ref{Figure:fig-3}. In 
Figure~\ref{Figure:fig-4}, we present an expanded one day light curve of 
the same source.  For 
most of the sources in our sample, observations are sparse in W3 and W4 bands 
having "null" entries at many epochs. Therefore, for any further analysis of 
variability, only two photometric bands were considered namely W1 and W2. Even 
when using W1 and W2 bands for variability analysis, to ensure use of good 
photometric measurements, following \cite{2019MNRAS.483.2362R}, the following 
conditions were imposed

\begin{enumerate}
\item The $\chi^{2}$ per degree of freedom of the single-exposure profile-fit in 
both W1 and W2 bands should be less than 5. 
\item  The number of components used in the fit of the point spread function (PSF) of
a source should be less than 3. 
\item The best quality single-exposure image frames are not affected by known artifacts and are not actively deblended. 
\item  The number of data points in a day must be at least 5 in both W1 and W2 bands. This condition was
utilized only for the analysis of variability on intra-day timescales. 
\end{enumerate}

\section{Analysis of Variability}
The conditions imposed in Section 2 
were used on the selection of data for variability analysis. For studying long term variability, we used
all such selected photometric points. However, for the analysis of variability on intra-day 
timescales we separated the photometric points into groups (hereafter called days). A one day
light curve includes all photometric points that have time gaps less than 1.2 days between
any two consecutive photometric points. Also, to get rid of cosmic rays affecting our photometric data
we employed a 3$\sigma$ clipping to remove outliers in the one day and long term 
light curves following  \cite{2019MNRAS.483.2362R}.
Thus, good quality photometric measurements were used in our analysis of mid infrared 
variability on intra-day timescales and long timescales.

\subsection{Variability amplitude}
To characterize the variability shown by a source both on  
intra-day and long timescales, we  calculated
the amplitude of variability ($\sigma_{m}$) as described in \cite{2007AJ....134.2236S}.  
For each light curve, $\sigma_{m}$ was calculated by removing the measurement uncertainty 
from the variance of the light curve.
According to \cite{2007AJ....134.2236S} $\sigma_{m}$ is given by
\begin{equation}
\sigma_{m} =
       \begin{cases}
            \sqrt{\Sigma^2 - \epsilon^2} & \text{if $\Sigma > \epsilon$,} \\
            0 & \text{otherwise}
        \end{cases}
\end{equation}
where $\Sigma$ is the variance of the light curve defined as
\begin{equation}
\Sigma = \sqrt{\frac{1}{n-1} \sum_{n=1}^{N} (m_i - <m>)^2}
\end{equation}
Here, $m_i$ is the magnitude of the i$^{th}$ point and $<m>$ is the weighted average. And, $\epsilon$ is
defined as  
\begin{equation}
\epsilon^{2} = \frac{1}{n} \sum_{n=1}^{N} \epsilon_i^{2}+\epsilon_s^{2}
\end{equation}
Here $\epsilon_i$ is the measurement uncertainty of the $i^{th}$ point and $\epsilon_s$ 
is the corresponding systematic uncertainty. For W1 and W2 bands, the systematic
uncertainties are 0.024 mag and 0.028 mag respectively \citep{2011ApJ...735..112J}. The systematic 
errors were added in quadrature to the measurement uncertainties to get the total
error on each photometric measurement.

\begin{figure}
\hspace*{-0.5cm}\includegraphics[scale=0.5]{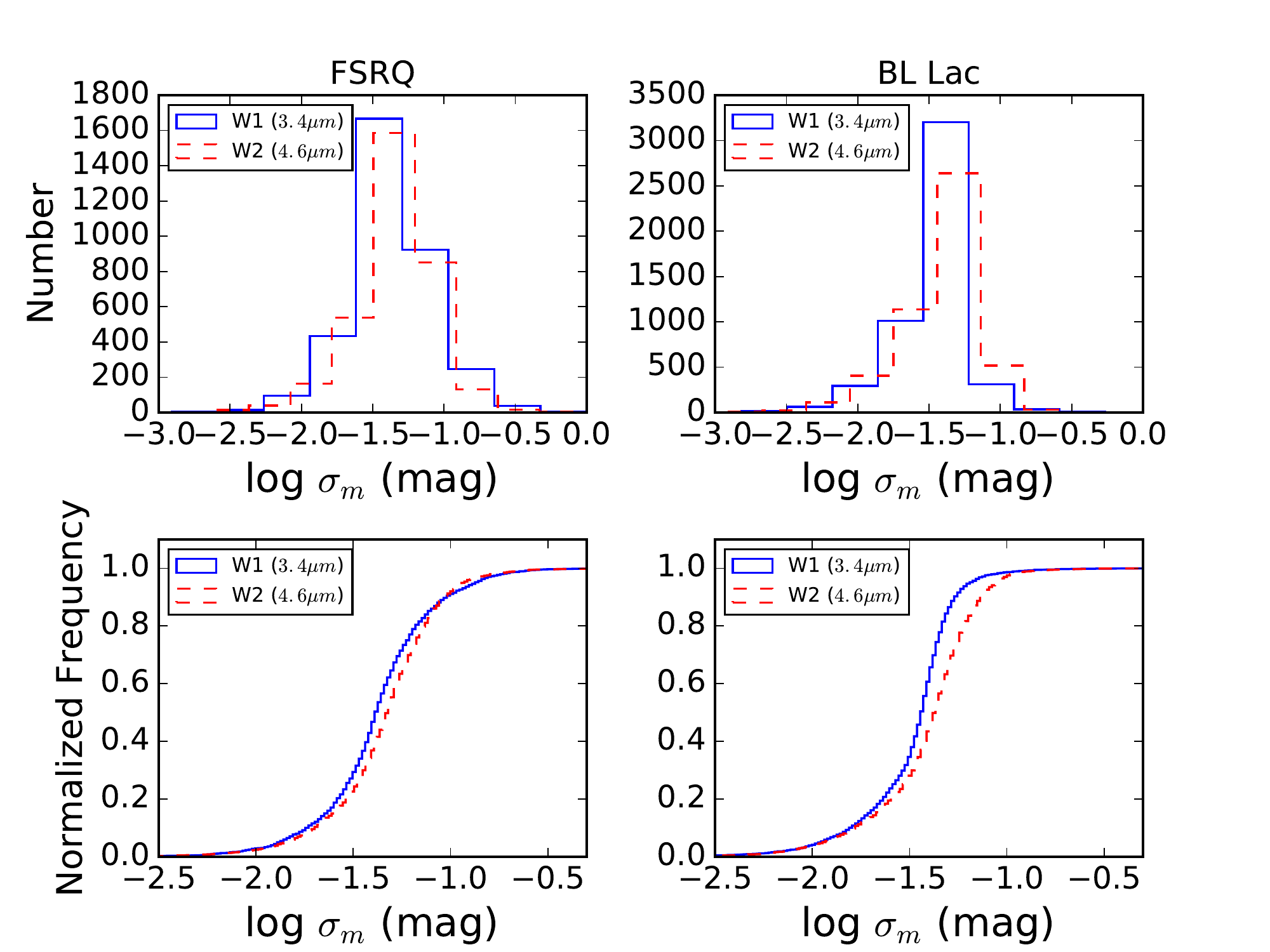} 
\caption{Histogram and cumulative distribution of $\sigma_m$ on intra-day timescales for
FSRQs (dashed line) and BL Lacs (solid line) in W1 (left panel) and W2 (right panel) bands.}
\label{Figure:fig-5}
\end{figure}

\subsubsection{Intra-day variability amplitude}
On intra-day times scales, the variability amplitude $\sigma_m$ in mag was
calculated using Equation 1. We found median $\sigma_m$ values of 0.04$^{+0.03}_{-0.02}$ mag and 0.05$^{+0.03}_{-0.02}$ mag for FSRQs in W1 and W2 bands. Similarly, for BL Lacs, we found median $\sigma_m$ values of 0.04$^{+0.01}_{-0.02}$ mag and 0.04$^{+0.02}_{-0.02}$  mag in W1 and W2 bands, respectively. The upper and lower uncertainties in the median
$\sigma_m$ values were determined such that 15.87\% of $\sigma_m$ values have $\sigma_m > \sigma_m(median) + \sigma_m(upper ~error)$ and 15.87\% of $\sigma_m$ values 
have $\sigma_m < \sigma_m(median) - \sigma_m(lower ~error)$. This corresponds to 1 $\sigma$ error
for a Gaussian distribution. 
The histogram and cumulative distribution of $\sigma_m$
for FSRQs and BL Lacs in W1 and W2 bands are shown in Figure \ref{Figure:fig-5}.  The median values
of $\sigma_m$ in W1 and W2 bands seem indistinguishable within errors for both FSRQs and BL Lacs. 
However, the two sample Kolmogorov-Smirnov (KS) test indicates that there is difference in the variability between
W1 and W2 bands on intra-day timescales. For FSRQs the KS test gave a D-value of 0.113, 
with a null hypothesis (there is no difference in variability between W1 and W2 bands) 
probability ($p$) of 2.14 $\times$ 10$^{-19}$, while for BL Lacs, from KS test we found a 
D-value of 0.213 with a $p$ of 8.79 $\times$ 10$^{-98}$.  Thus, on intra-day timescales
there is difference in the mean variability amplitude between W1 and W2 bands in both FSRQs and BL Lacs.
From our analysis we found FSRQs to show similar median amplitude of variability to BL Lacs in both W1 
and W2 mid-IR bands. However, a two sample KS test carried out for the distribution of 
$\sigma_m$ in the W1 band in FSRQs and BL Lacs showed that the two distributions are indeed
different with a D-statistic of 0.211 and a null-hypothesis (the distribution
of $\sigma_m$ in W1 band for FSRQs and BL Lacs are drawn from the same population) 
 $p$ of 3.60 $\times$ 10$^{-79}$. Similarly in W2 band too, from KS test we 
found that the distribution of $\sigma_m$ are different between FSRQs and BL Lacs
with a D-statistic value of 0.121 and a p-value of 
6.86 $\times$ 10$^{-26}$. A summary of the results on intra-day variability analysis is included in Table \ref{Table:table-2}.

\begin{figure}
\hspace*{-0.5cm}\includegraphics[scale=0.5]{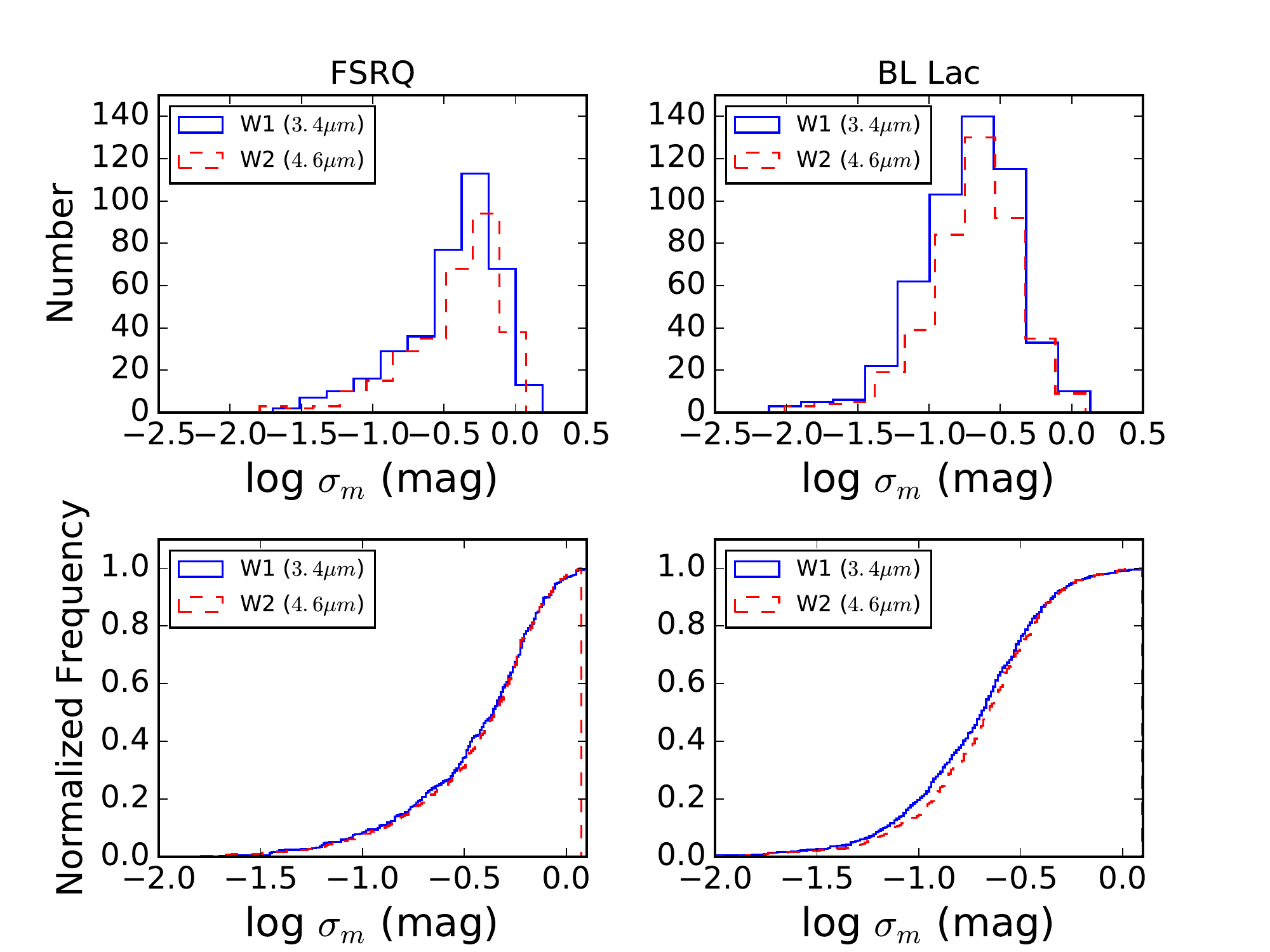} 
\caption{Histogram and cumulative distribution of $\sigma_m$ on long timescales for 
FSRQs (dashed line) and BL Lacs (solid line) in both W1 (left panel) and W2 (right panel) bands.}
\label{Figure:fig-6}
\end{figure}

\subsubsection{Long term variability amplitude}
The distribution of long term variability amplitudes in W1 and W2 bands
for FSRQs and BL Lacs are shown in Figure \ref{Figure:fig-6}. For FSRQs in W1 band we found $\sigma_m$ to
range between 0.020 mag and 1.550 mag with a median value of 0.44$^{+0.28}_{-0.27}$ mag. Similarly
for W2 band we found $\sigma_m$ to range between 0.016 mag and 1.186 mag with a median value of 0.45$^{+0.27}_{-0.27}$ mag. From two sample KS test applied to the distribution 
of $\sigma_m$ between W1 and W2 band, we found a D-statistic of 0.044 with a p-value of
0.90. Thus in FSRQs there is no difference in variability between W1 and W2 bands.
In the case of BL Lacs, in W1, we found $\sigma_m$ to range between 0.008 mag and 1.350 mag with
a median of 0.21$^{+0.18}_{-0.12}$ mag. Similarly in W2 band, we found $\sigma_m$ to 
lie in the range between 0.010 mag and 1.249 mag with a median of 0.22$^{+0.18}_{-0.11}$ mag. A two sample KS test to the
distribution of $\sigma_m$ between W1 and W2 bands in BL Lacs gave a D-statistic of 
0.070 with a p-value of 0.210. Thus in long term the amplitude of flux variations between W1 and
W2 bands are found to be similar in both FSRQs and BL Lacs. A two sample KS test
applied to the distribution of $\sigma_m$ in W1(W2) bands between FSRQs and BL Lacs yielded
a D-statistics of 0.42 (0.43) and a p-value of 1.10 $\times$ 10$^{-33}$ (2.23 $\times$ 
10$^{-28}$). Thus, on long timescales too, FSRQs showed larger amplitude variations than BL Lacs in 
both W1 and W2 bands. A summary of the results of the long term variability 
analysis is included in Table \ref{Table:table-2}. 

\subsection{Flux variability on sub-samples of blazars}
\begin{figure}
\hspace*{-0.5cm}\includegraphics[scale=0.5]{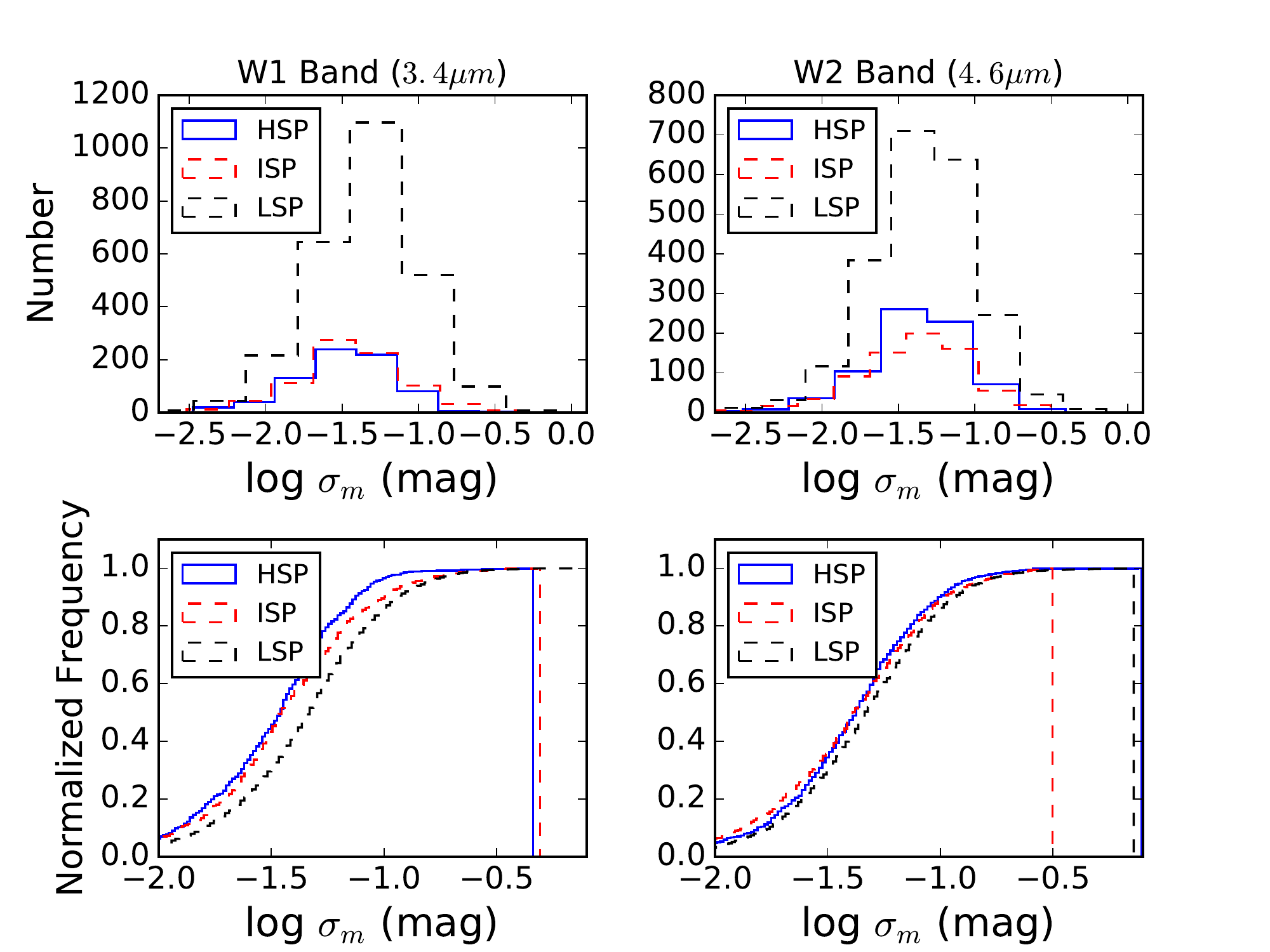} 
\caption{Distribution of $\sigma_m$ on intra-day timescales  
for LSP, ISP and HSP objects.} 
\label{Figure:fig-7}
\end{figure}

\begin{figure}
\hspace*{-0.5cm}\includegraphics[scale=0.5]{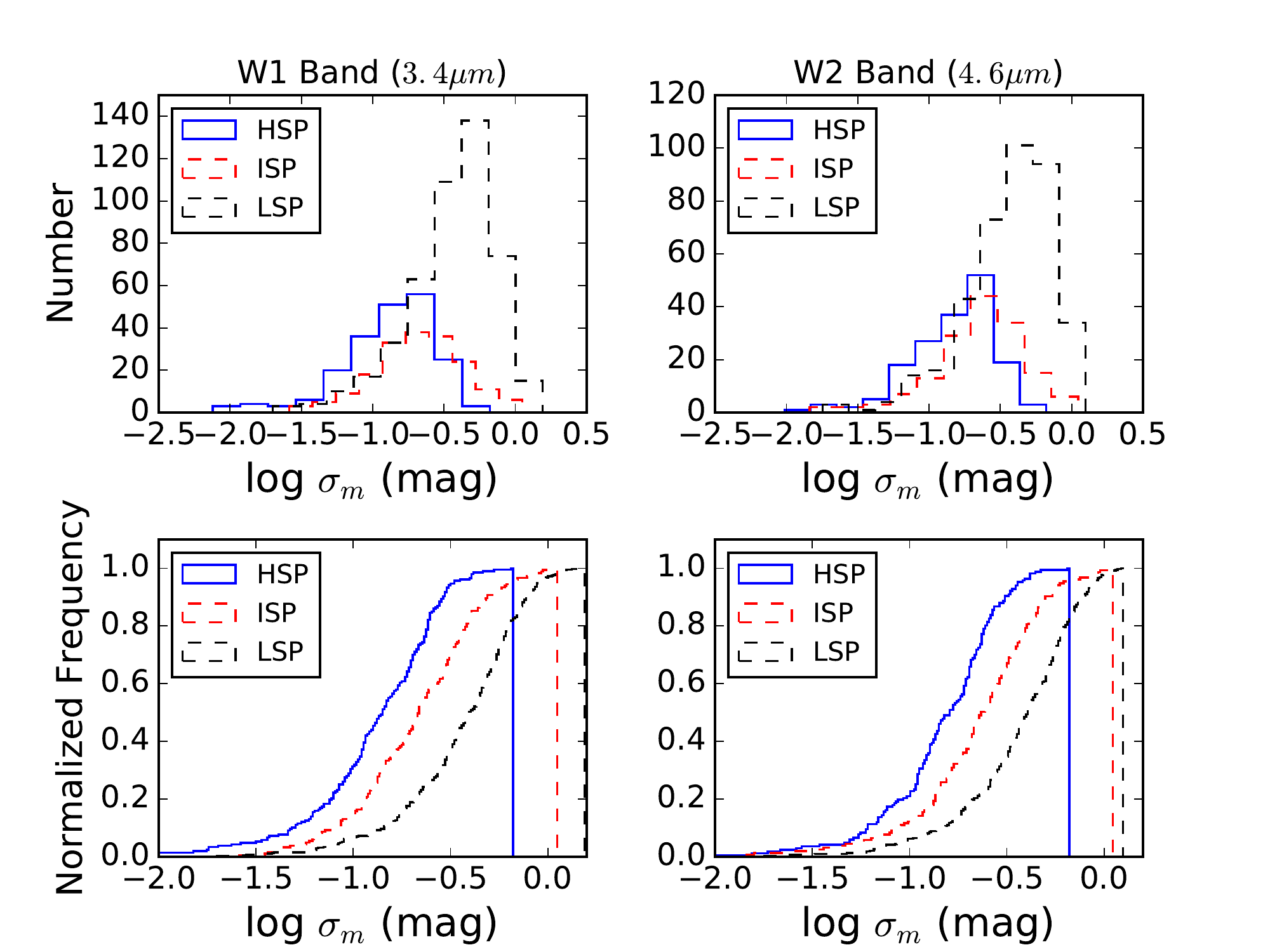}
\caption{Histogram and cumulative distribution of $\sigma_m$ on long timescales for the
different sub-classes of blazars in both W1 and W2 bands.}
\label{Figure:fig-8}
\end{figure}

We also divided our sample of blazars into different subclasses based on the peak frequency
of the synchrotron component in their broad band SED such as LSP, ISP and HSP and analysed
the amplitude of variability in them both on intra-day and long timescales.
The results are included in Table \ref{Table:table-2} for both intra-day and long timescales respectively.
The histogram and cumulative distribution of $\sigma_m$ for different sub-classes
of blazars are given in Figure \ref{Figure:fig-7} for intra-day timescales and 
Figure \ref{Figure:fig-8} for long timescales.
On long timescales LSP blazars showed the largest $\sigma_m$ followed by ISP and HSP sources in both
W1 and W2 bands. This is evident from the cumulative distribution function (CDF) plots in the bottom panels of Figure \ref{Figure:fig-8}, where the CDFs of LSPs are systematically at higher $\sigma_m$ values than those of ISPs and HSPs, while the CDFs of ISPs are larger than HSPs.
This is expected from the results in the previous section as majority of LSP sources are FSRQs while most of HSP sources are BL Lacs.
On intra-day times scales, while LSP sources are more variable compared to ISP and HSP sources, the 
amplitude of variability is indistinguishable between ISP and HSP sources. 
 Here too, in the CDFs shown in the bottom panels of Figure \ref{Figure:fig-7}, LSPs have higher $\sigma_m$ values than ISPs and HSPs, while the CDFs of ISPs and HSPs are indistinguishable.

\subsection{Ensemble Structure Function}
In addition to the variability of blazars characterized by $\sigma_m$ described in
the previous sections, blazar variability  can also be described by Structure function (SF), which 
shows the dependency of variability as a function of time lag 
\citep{1985ApJ...296...46S}. SF can be calculated for individual AGN having 
light curve with multiple epochs of observations, taking the magnitude difference 
for each pair of time lags in a light curve. It can also be calculated for a 
group of blazars, known as ensemble structure function, allowing us to obtain the 
mean variability behaviours of the population. Since WISE data has sparse sampling, 
we studied the mean variability of different classes of blazars using the ensemble 
structure function following \cite{1996ApJ...463..466D} 

\begin{equation}
\mathrm{SF} = \sqrt{\frac{\pi}{2} <|\Delta m|>^2 - <\sigma^2_n>},
\end{equation}     
where $|\Delta m|=m_i - m_j$, is the magnitude difference of any two epochs 
($i$, $j$) separated by the time difference or lag $\Delta \tau = t_i - t_j$. 
$\sigma^2_n=\sigma^2_i+\sigma^2_j$ is the sum of squares of the uncertainties  
in magnitudes m$_i$ and m$_j$. To calculate SF, the time bins were
selected to have equal interval in 
logarithmic observer frame time lag with at least 2000
data points in each bin. Note that a few of our sources do not have redshift 
measurements in the literature, thus, SF was calculated in the observer frame.

\begin{figure}
\includegraphics[scale=0.6]{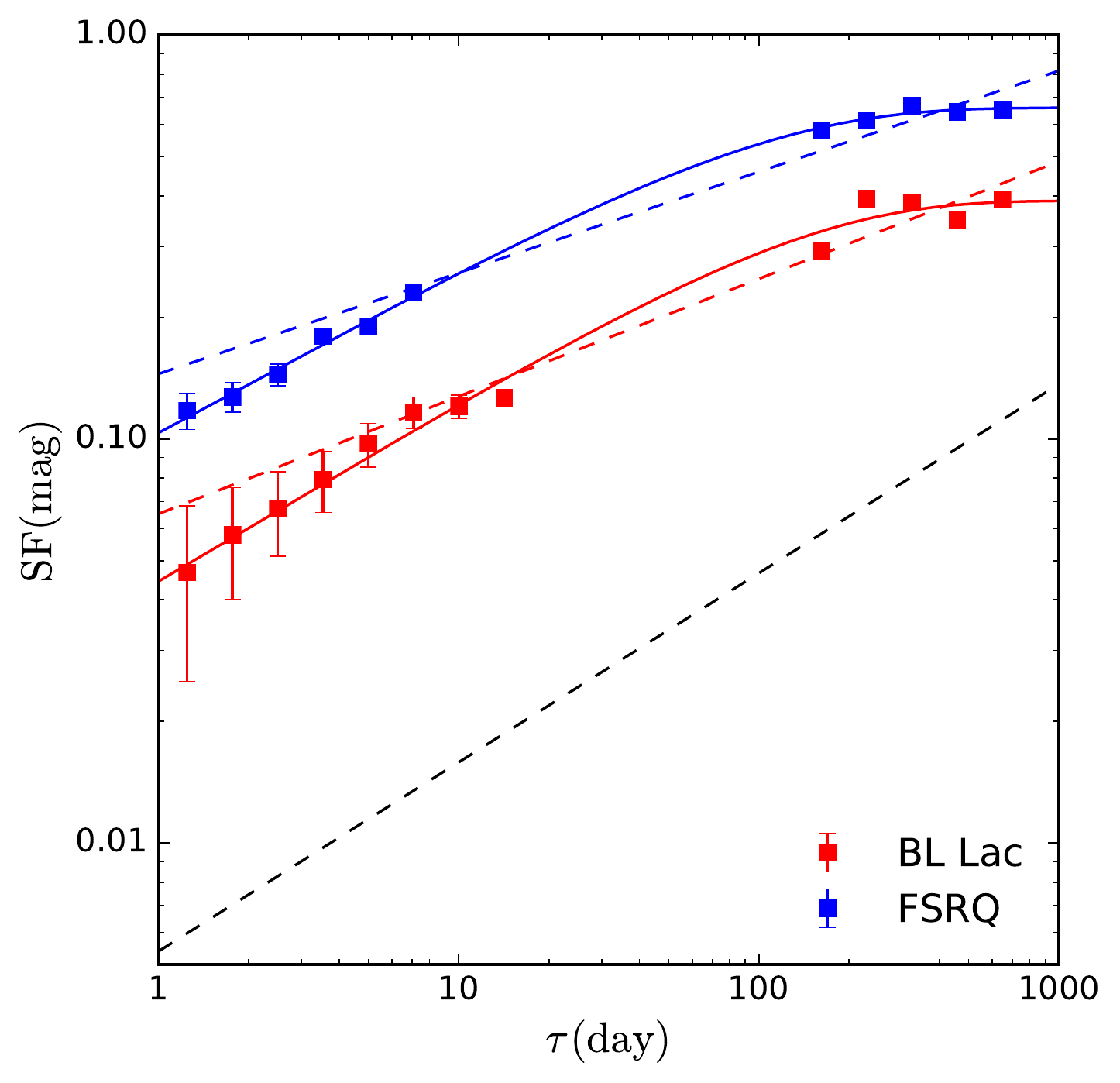}
\caption{Structure function (SF) against observer frame time lag for BL Lacs (red dots) and FSRQs (blue dots). Best fits of the SF using Equation 2 (dashed line) and Equation 3 (solid-line) are also shown. The 3.6 micron structure function for quasars from \citet{2016ApJ...817..119K} is shown with a black dashed line.}
\label{Figure:fig-9} 
\end{figure}

In Figure \ref{Figure:fig-9}, we plot SF against observer frame time lag 
for BL Lacs (red) and FSRQs (blue). The error bars in SF were calculated via error 
propagation following \cite{2004ApJ...601..692V}. The Figure shows 
FSRQs are significantly more variable than BL Lacs consistent with the result obtained by 
$\sigma_m$ calculation. SF increases gradually from a time lag of  1 to $\sim200$ 
days and become flatter at higher time lag. Such a trend has been noted previously 
by various authors (e.g., \citealt{2004ApJ...601..692V,2011A&A...527A..15W, 2016ApJ...826..118K})
from studies of long term quasar variability in the optical band. 
To characterize the structure function, we fitted a simple power-law model in the form of 
       
\begin{equation}
\mathrm{SF} = \left(\frac{\Delta \tau}{\tau_0}\right)^{\gamma},
\label{eq:SF_PL}
\end{equation}     
where $\tau_0$ and the power law slope ($\gamma$) are free parameters. The fitting results are given 
in Table \ref{Table:table-3}. The inferred value of $\gamma$ is 
$0.29 \pm 0.03$ for BL Lacs and $0.25 \pm 0.02$  for FSRQs. We also fitted a 
three parameters exponential model following \cite{2016ApJ...826..118K} in the form of 
\begin{equation}
\mathrm{SF} = \mathrm{SF_{\infty}} \sqrt{ 1- e^{- \left(\frac{|\Delta \tau|}{\tau_c} \right)^{\beta}}},
\label{eq:SF_EXP}
\end{equation}  
 where $\mathrm{SF_{\infty}}$ is the amplitude of variance at long timescale,  
$\tau_c$ is de-correlation timescale and $\beta$ is the power-law index. The 
equation becomes a damped random walk (DRW;\citealt{2009ApJ...698..895K}) 
 for $\beta=1$. We note that the de-correlation timescales could be 
affected by the limited length of the light curves as shown from simulations
by \cite{2010MNRAS.404..931E}.
 
\begin{figure}
\includegraphics[scale=0.6]{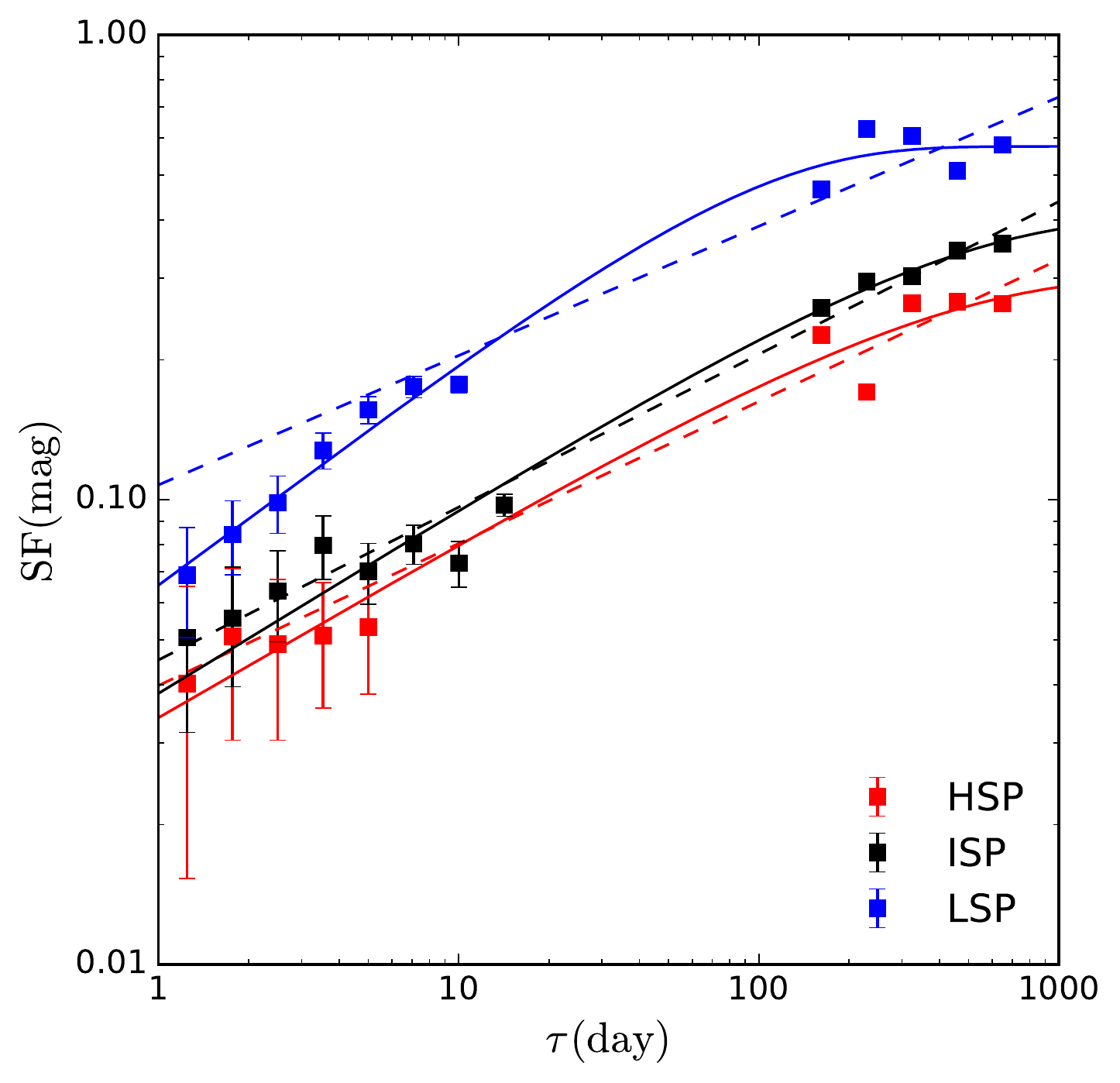}
\caption{Structure functions for HSP (red), ISP (black) and LSP (blue). Lines 
have the same meaning as in Figure \ref{Figure:fig-9}.}
\label{Figure:fig-10}. 
\end{figure}

Equation \ref{eq:SF_EXP} provides a better fit to the data than simple 
power-law defined in Equation \ref{eq:SF_PL} especially at lower lag. The fitted 
parameters 
are given in Table \ref{Table:table-3}. We found $\mathrm{SF_{\infty}}=0.38 
\pm 0.02$ for BL Lacs and $0.66 \pm 0.01$ for FSRQs suggesting a higher variability 
in FSRQs than BL Lacs. The value of $\beta$ is about $\sim0.85$ slightly deviating from 
DRW. We also plot in the same figure, the 3.6 micron SF of confirmed quasars from 
\cite{2016ApJ...817..119K} in black dashed line, which lies well below   
that of the BL Lacs and FSRQs studied here. This could be due to the contribution 
of relativistic jets to the mid infrared variability of the blazars studied
here in comparison to the sample of quasars studied by \cite{2016ApJ...817..119K}.
In Figure \ref{Figure:fig-10}, we plot the SFs of HSP, ISP and LSP sources. Here too, we 
found a better fit with the exponential function given in  Equation \ref{eq:SF_EXP}. 
We found $\mathrm{SF_{\infty}}=0.30 \pm 0.07$, $0.40 \pm 0.04$ and $0.57 \pm 0.03$ 
for HSPs, ISPs, and LSPs. 

 The SF plots in Figure \ref{Figure:fig-10} show that variability
is significantly stronger in LSPs than in ISPs and variability is the lowest in HSPs. Among ISPs and HSPs, 
variability is stronger in ISPs. This is in agreement with the results obtained
from the analysis of the amplitude of flux variations.
 
\begin{table}
\caption{Details of the sources used in this study.}
\begin{center}
\resizebox{0.9\linewidth}{!}{%
\begin{tabular}{ l l l c }\hline  
Type    &  Number   & $z$ range  & Median $z$ \\ \hline  
    BL Lac     & 460 & 0.189$-$3.10  & 1.106 \\ 
    FSRQ       & 575 & 0.034$-$1.72  & 0.291 \\
    HSP	       & 565 & 0.085$-$1.25  & 0.770 \\	
    ISP	       & 207 & 0.046$-$2.19  & 0.046 \\	
    LSP	       & 243 & 0.034$-$1.60  & 0.203 \\ \hline	
\end{tabular} } 
\label{Table:table-1}
\end{center}
\end{table}

\begin{table*}
\caption{Results of the analysis of variability.}
\label{Table:table-2}
\resizebox{0.9\linewidth}{!}{%
\begin{tabular}{ | l | c | c | c | c | c | c | c | } \hline  
Type  & Number & \multicolumn{2}{c}{$\sigma_{m} \pm ~\sigma$} (intraday timescale) & \multicolumn{2}{c}{$\sigma_{m} \pm ~\sigma$ (long timescale)}  & \multicolumn{2}{c}{Duty cycle } \\ \cline{3-8}
      &        &   W1(mag)   &   W2(mag)  & W1(mag)   & W2(mag)  &  W1(\%)   & W2(\%)  \\  \hline
FSRQ  & 460  & 0.04$^{+0.03}_{-0.02}$ & 0.05$^{+0.03}_{-0.02}$ & 0.44$^{+0.28}_{-0.27}$ & 0.45$^{+0.27}_{-0.27}$ & 78.82 & 53.06 \\ 
BL Lac & 575 & 0.04$^{+0.01}_{-0.02}$ & 0.04$^{+0.02}_{-0.02}$ & 0.21$^{+0.18}_{-0.12}$ & 0.22$^{+0.18}_{-0.11}$ & 89.59 & 61.47 \\
LSP   & 565  & 0.05$^{+0.05}_{-0.03}$ & 0.05$^{+0.05}_{-0.03}$ & 0.40$^{+0.30}_{-0.22}$ & 0.41$^{+0.29}_{-0.22}$ & 79.51 & 54.12 \\ 
ISP   & 207  & 0.04$^{+0.04}_{-0.02}$ & 0.04$^{+0.05}_{-0.02}$ & 0.22$^{+0.19}_{-0.11}$ & 0.24$^{+0.21}_{-0.12}$ & 92.79 & 82.29 \\
HSP   & 243  & 0.04$^{+0.03}_{-0.02}$ & 0.04$^{+0.04}_{-0.02}$ & 0.14$^{+0.11}_{-0.08}$ & 0.16$^{+0.10}_{-0.09}$ & 82.89 & 43.71 \\ \hline 
\end{tabular} } 
\end{table*}

\begin{table*}
\caption{Results of model fits to the ensemble structure function for various classes of blazars.}
\begin{center}
\resizebox{0.9\linewidth}{!}{%
\begin{tabular}{ l l l c c c}\hline  
                & \multicolumn{2}{c}{Power law model}  & \multicolumn{3}{c}{Exponential model} \\ \cline{2-6}
Object class    &   $\gamma$        &  $\tau_0$ (days) & $\mathrm{SF_{\infty}}$ (mag) & $\beta$ &  $\tau_c$ (days)         \\ \hline
    BL Lac      & $0.29 \pm 0.03$   & $1.19 \pm 0.52 \times 10^4 $  & $0.38 \pm 0.02$   &  $0.89 \pm 0.09$ & $128 \pm 46$  \\		
    FSRQ        & $0.25 \pm 0.02$   & $2.25 \pm 0.63 \times 10^3 $  & $0.66 \pm 0.01$   &  $0.82 \pm 0.03$ & $90 \pm 12$   \\
    HSP		& $0.30 \pm 0.05$   & $3.81 \pm 3.26 \times 10^4 $  & $0.30 \pm 0.07$   &  $0.75 \pm 0.14$ & $327\pm 410$  \\
    ISP		& $0.32 \pm 0.02$   & $1.23 \pm 0.28 \times 10^4 $  & $0.40 \pm 0.04$   &  $0.79 \pm 0.05$ & $375\pm 167$  \\
    LSP		& $0.27 \pm 0.04$   & $3.01 \pm 1.29 \times 10^3 $   & $0.57 \pm 0.03$   &  $0.96 \pm 0.13$ & $89\pm 39$   \\  \hline
\end{tabular} } 
\label{Table:table-3}
\end{center}
\end{table*}
       
\subsection{Duty cycle of variability}
The data set analysed here for intra- day variability is suited to estimate the 
duty cycle (DC) of mid infrared variability of different classes of blazars. 
Following \cite{1999A&AS..135..477R}, the DC of a particular class
of object is given as
\begin{equation}
DC =  100 \frac{\sum_{j=1}^{n} A_j \times (1/\Delta t_j)}{\sum_{j=1}^{n} (1/\Delta t_j)} \%
\end{equation}
Here, $\Delta t_j = \Delta t_{j, obs}/(1+z)$ is the duration of observation corrected for the 
cosmological redshift of the source of a $j^{th}$ one day light curve (as 
described in Section 3) for a selected class. $A_j$ = 1 or 0, 
if the source was classified as a variable or non-variable respectively during the observed duration $\Delta t_j$.
 A source is considered variable in a one day light curve ($A_j$ = 1) if $\sigma_m$ calculated
using Equation 1 is greater than zero.
For BL Lacs in the W1 band we found a high DC of 89.59\%, while for FSRQs we 
found a lower DC of variability of 78.82\%. In the W2 band too, 
BL Lacs have a high DC of 61.47\% followed by FSRQs of 53.06\%.   Separating
the blazars into different classes based on their SED, we found  the ISP sources to have the highest DC of variability 
and HSPs have the lowest DC of variability in both W1 and W2 bands. The results of the DC of variability 
in various sub-classes of blazars is given in Table \ref{Table:table-2}. 
In the optical band on intra-night timescales, \cite{2004JApA...25....1S} found the BL Lacs to have a 
high DC of variability related to FSRQs which was interpreted as due to the close alignment of BL Lac jets with the line of sight related to FSRQs \citep{2004JApA...25....1S}.  The results found here on DC of  the mid infrared variability 
on intra-day timescales closely matches with that found by \cite{2004JApA...25....1S}  on intra-night
timescales in the optical band in spite of differences in the time resolution between the optical and the 
mid infrared light curves. 

\subsection{Colour Variability}

Blazars are known to show spectral variations in the optical band. It 
has been thought that FSRQs show a redder when brighter behaviour 
(RWB; \citealt{2006A&A...450...39G,2012ApJ...756...13B}). Alternatively, BL Lacs are 
found to show a bluer when brighter behaviour (BWB; \citealt{1998MNRAS.299...47M,
2002A&A...390..407V,2003ApJ...590..123V,2012MNRAS.425.3002G}). Departures from 
this conventional observations have also been noted recently. FSRQs are found 
to show bluer when brighter behaviour \citep{2011A&A...528A..95G} and in the
FSRQ 3C 345 both RWB and BWB trends were noticed \citep{2011MNRAS.418.1640W}. There 
are also reports in which the spectrum of a blazar was 
found not to change with increasing/decreasing 
brightness \citep{2006MNRAS.366.1337S}. 

 The available literature is more focussed on the colour variability of blazars in the optical
and near infrared bands \citep{2019MNRAS.484.5633G,2019ApJ...887..185S,2012ApJ...756...13B}, but not in the
mid infrared bands.
Most of these studies were based on 
nearly quasi-simultaneous observations, without properly  taking into
account the errors in both colours and magnitudes which
could lead to incorrect characterisation of spectral variability 
\citep{2016RAA....16...27S}. In this work we report on the mid infrared spectral variations in a large sample of blazars.

\subsubsection{Intra-day colour changes}

\begin{figure}
\hspace*{-0.5cm}\includegraphics[scale=0.5]{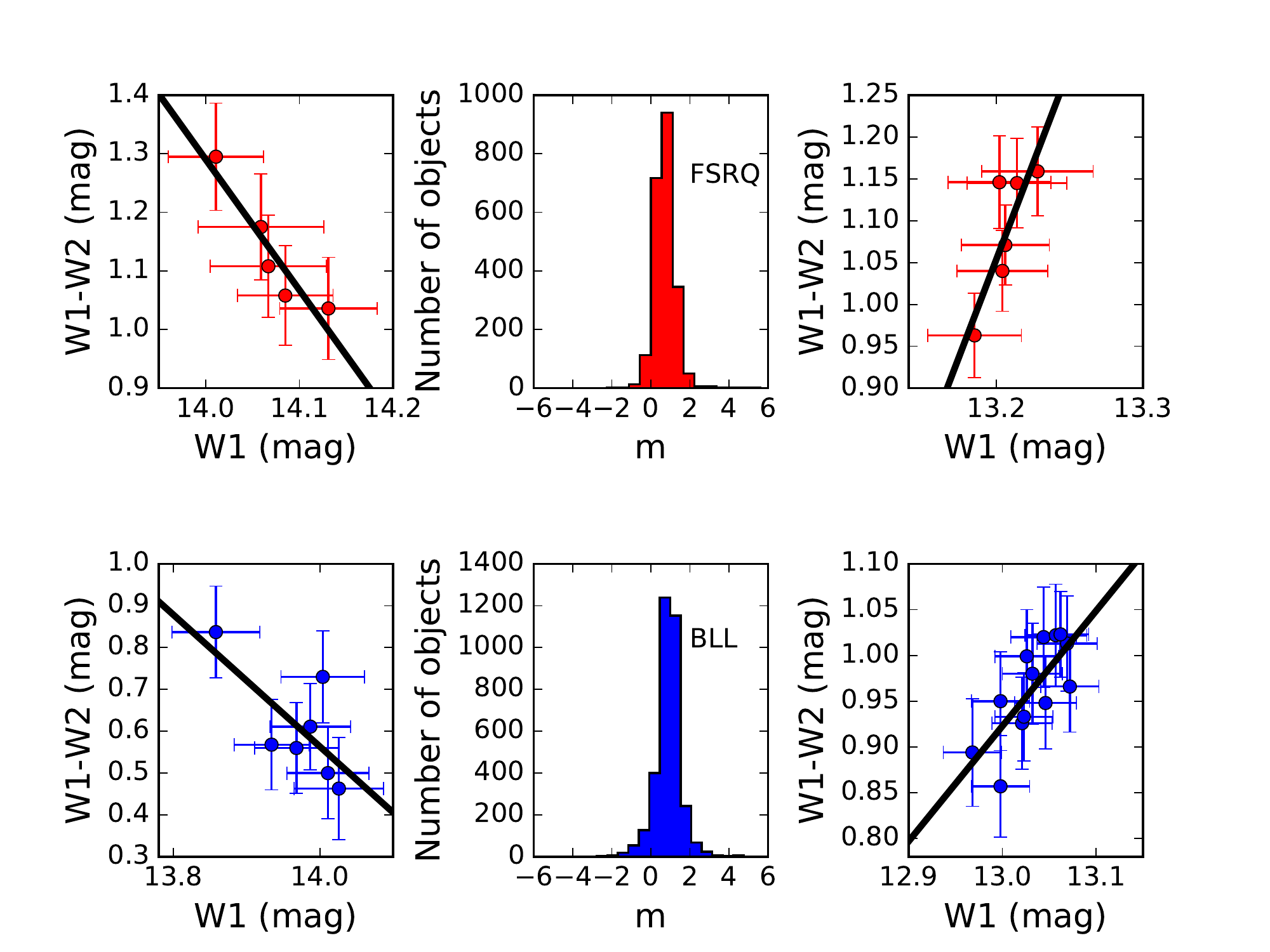}
\caption{Distribution of the slopes of the linear least squares fits done to the colour magnitude
diagram of FSRQs (top middle panel) and BL Lacs (bottom middle panel) on intra-day timescales.
The left and right panels on the top show a typical example of a RWB and BWB trend, respectively in FSRQs, while
the bottom left and bottom right panels show a sample RWB and BWB trend in BL Lacs, respectively}.
\label{Figure:fig-11}
\end{figure}

The advantage of colour variability studies in mid infrared using WISE is that 
the observations in the different bands are simultaneous. To characterise the colour variability in W1 
and W2 bands, we constructed colour - magnitude diagrams, wherein (W1-W2) colour is plotted along the Y-axis and the W1 brightness is plotted along 
the X-axis.  
We carried out a linear least squares                                        
fit to the colour - magnitude diagram by taking into account the errors 
in both the colour and magnitude. The
slope of the fit is taken to quantify the spectral change. We used the Spearman rank correlation analysis to probe the correlation between W1-W2 colour against the W1 brightness. The source becomes increasing bluer with increasing W1-W2 colour along the Y-axis and W1 increasing (decreasing in brightness) towards the right. This is the BWB trend as seen in Figure \ref{Figure:fig-11} (right panel). Alternatively, the situation wherein the W1-W2 colour gets smaller with increasing W1 band (decreasing in brightness) is the RWB trend. An example of such a trend is shown in the left panel of Figure \ref{Figure:fig-11}. In this work we adopted the following criteria to characterize the colour variations in blazars. We considered a source to show a BWB trend if the Spearman rank correlation coefficient is larger than 0.3 and probability of no correlation (p) is less than 0.05. Similarly we considered a source to show a RWB trend if the Spearman rank correlation coefficient is less than $-$0.3 and $p$ is less than 0.05.
In the middle top panel of Figure \ref{Figure:fig-11}  is shown the distribution of
the slopes obtained from linear least squares fit to the colour magnitude 
diagram on intra-day timescales. Clearly the blazars in our sample showed
all types of spectral behaviours namely (a) constancy of spectral shape with brightness, (b) RWB behaviour 
 and (c) BWB behaviour. However, the distribution is shifted from zero to positive 
 values (there are more positive than negative values)
thereby indicating that on intra-day times scales most of the blazars
showed a BWB trend. On intra-day timescales, in BL Lacs, for 1409 light curves, we
found a BWB trend, while only on four light curves, we noticed a RWB trend. For FSRQs,
960 light curves showed a BWB trend, while only on five light curves RWB trend was noticed.
Thus, on intra-day timescales we found
both FSRQs and BL Lacs predominantly showed a BWB trend, while only on very few instances
RWB trend was noticed.

\begin{figure}
\hspace*{-0.5cm}\includegraphics[scale=0.50]{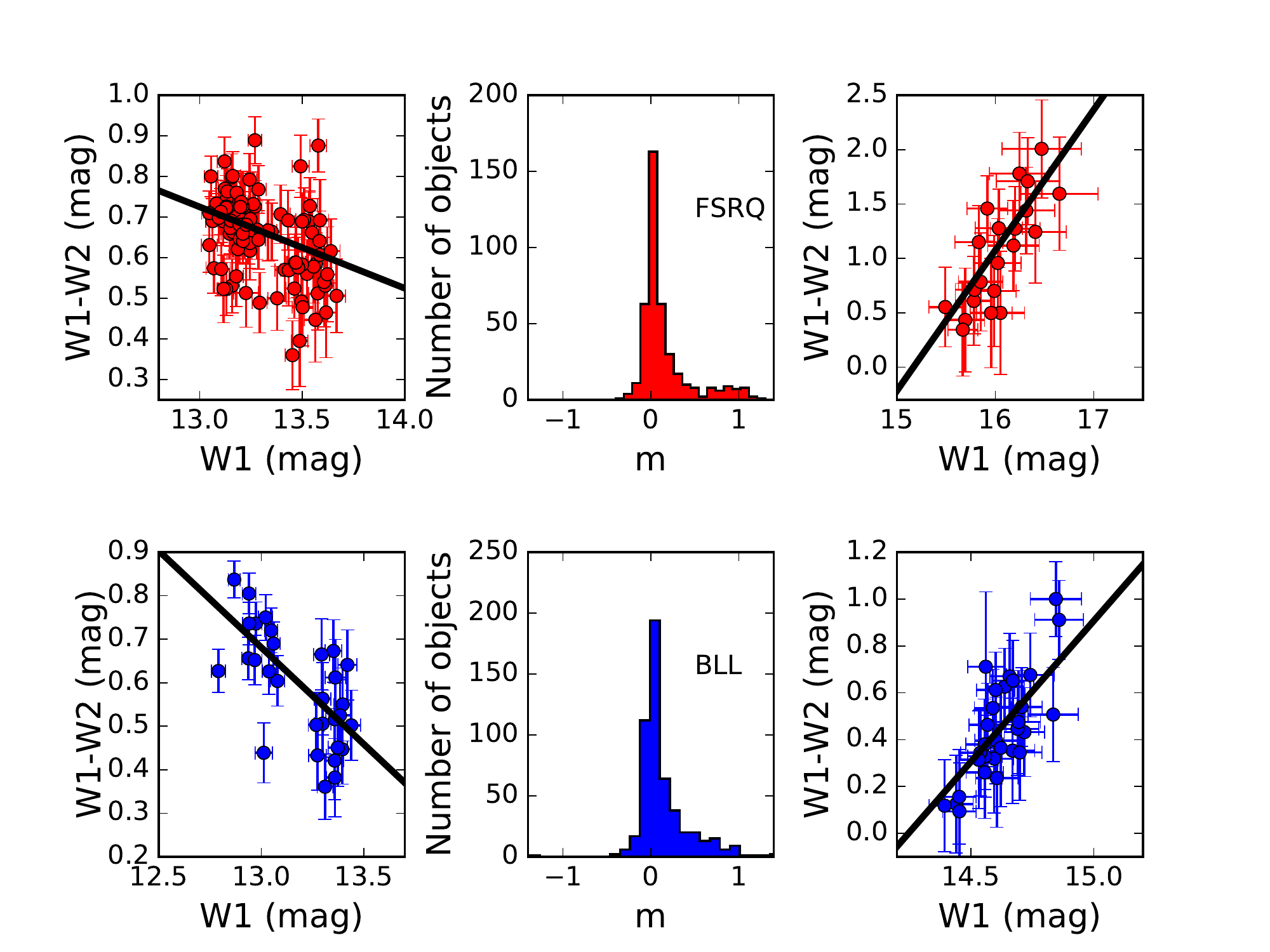}
\caption{The top middle panel and the bottom middle panel shows the distribution of slopes obtained
from the linear least squares fit to the colour magnitude diagram on long timescales for FSRQs and
BL Lacs. The top left and top right panels show sample RWB and BWB trend, respectively in FSRQs, while
the same for BL Lacs is shown in the bottom left and bottom right panels, respectively}.
\label{Figure:fig-12}
\end{figure}

\subsubsection{Long term colour changes}

Using data that spans from about a year to as long as 7 years, we also studied 
the colour changes in the long timescales. Here too, colour magnitude diagrams were 
constructed, and linear least square fits were carried out to the colour-magnitude
diagrams by taking into account the errors in both the colours and magnitudes. The results of the 
spectral fits are shown in Figure \ref{Figure:fig-12}. Similar to the spectral trends noticed
on intra-day light curves, here too, we found all types of spectral behaviours 
namely, (a) the spectral shape has not changed with 
brightness (b) the spectrum has become BWB and (c) the spectrum has become 
RWB. An example of a BWB character is shown in the bottom right panel of 
Figure \ref{Figure:fig-12}, while an example of a RWB trend is shown on the bottom left hand panel of Figure \ref{Figure:fig-12}.
The distribution of the slopes of the spectral fits are shown in the middle panels of
Figure \ref{Figure:fig-12}. Similar to the spectral variations observed from intra-day light curves, here too, 
the distribution of spectral slopes have more positive than negative values, pointing to the 
BWB trend in most of the blazars. On long timescales, we found 174
FSRQs to show a BWB trend, while RWB trend was found in 24 sources. Similarly, for BL Lacs, 160 objects showed a BWB trend, and a small number of 47 objects showed a RWB trend. Thus, on long timescales majority of FSRQs and BL Lacs showed a BWB spectral behaviour.

\begin{figure}
\hspace*{-0.5cm}\includegraphics[scale=0.50]{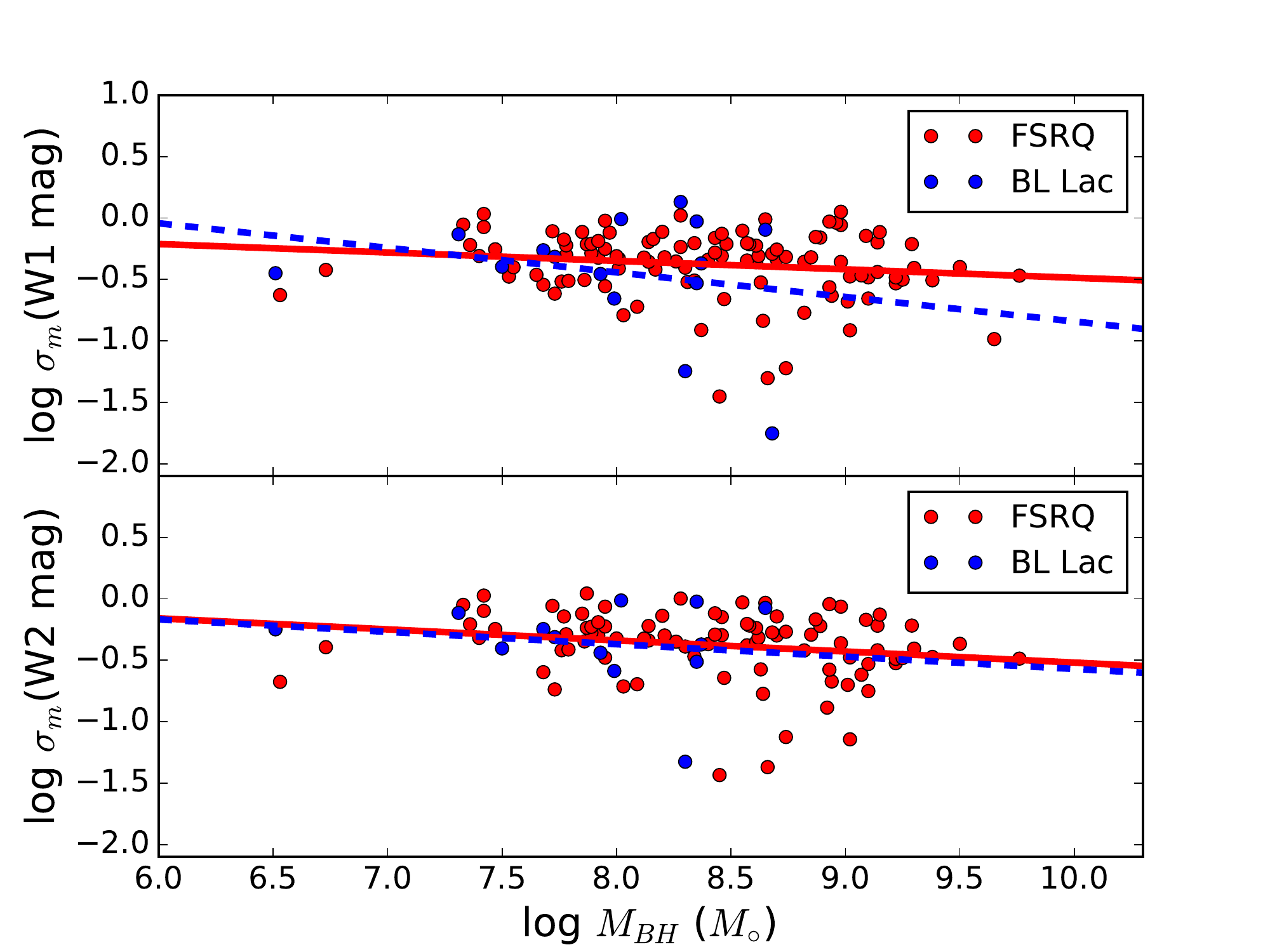}
\caption{Correlation of $\sigma_m$ with M$_{BH}$. The solid and dashed lines
are unweighted linear least squares fit to the data on FSRQs and BL Lacs respectively.}
\label{Figure:fig-13}
\end{figure}

\subsection{Correlation of long timescale $\sigma_m$ with $M_{BH}$}
In order to find the correlation between the black hole mass ($M_{BH}$)    
 and long timescale $\sigma_m$, we collected $M_{BH}$ values from the catalog of virial black hole 
mass estimates obtained by \citet{2016ApJ...817..119K}. We could get $M_{BH}$ values 
for a total of 67 objects. For those objects, we checked for the correlation between 
$M_{BH}$ and $\sigma_m$ in both W1 and W2 bands, by 
 applying a linear least square fit to the logarithmic variability amplitude $\sigma_m$ as a function of the logarithmic black hole mass.
This is shown in Figure \ref{Figure:fig-13}. 

For FSRQs in the W1 band, the Spearman rank correlation coefficient is $-$0.15 with a $p$ value of 
0.11 and for the W2 band, the correlation coefficient was found to be $-$0.20 
with a $p$-value of 0.06. For BL Lacs we found the correlation coefficient of $-$0.23 with a $p$ of 
0.41 in W1 band, while for W2 band we found a correlation coefficient and $p$ of $-$0.17 and 0.59 
respectively. Thus, in our data we did not find statistically significant correlation between
$M_{BH}$ and mid infrared amplitude of variability. 
This is the first investigation of the correlation of
$\sigma_m$ with $M_{BH}$ for blazars in the mid infrared bands, however, such  correlation analysis  
are already available in other wavelengths in other types of AGN. For 
example in the optical band for quasars, there is no conclusive
evidence on the dependence of variability amplitude with  M$_{BH}$ as there are reports of 
positive \citep{2007MNRAS.375..989W,2019ApJ...877...23L},negative \citep{2009ApJ...698..895K} 
and no correlation \citep{2016A&A...585A.129S} 
on timescales similar to the one probed here in the mid infrared band. \cite{2019MNRAS.483.2362R} studied mid infrared variability of Narrow Line Seyfert 1 galaxies and
found no correlation between variability amplitude and M$_{BH}$.

\section{Discussion}
\subsection{Flux Variability}
The observations from WISE, that makes about 15 orbits per day is well suited to study the mid infrared flux
variations on intra-day timescales as well as long timescales. We utilised this data set to probe the
mid infrared variability characteristics of a sample of blazars selected from 3FGL. 
Our analysis is a first characterisation of the mid infrared variability of $\gamma$-ray emitting blazars. We found
that on intra-day timescales FSRQs showed larger amplitude variability than 
BL Lacs. In the optical band on timescales much shorter
than a day, BL Lac objects were found to show large amplitude and high duty cycle of variability compared to
FSRQs. These observations in the optical band are explained in the context of BL Lacs having closely aligned
jets relative to FSRQs \citep{2004JApA...25....1S}. The differences between the optical and mid infrared intra-day variability characteristics of blazars could be ascribed to the following: (i) the optical analysis of \cite{2004JApA...25....1S} was based on a limited number of 15 blazars, with each blazar having three or four intra-day light curves. The mid infrared variability obtained here is based on the analysis of a larger number of blazars with each blazar having many ($>$ 3) intra-day light curves and (ii) the time resolution of the optical light curves analysed by \cite{2004JApA...25....1S} is of the order of minutes, while, the time resolution of the mid infrared intra-day light curves analysed here is of the order of hours. Mid infrared light curves with time resolution of the order of minutes are needed to make a direct comparison to the results reported by \cite{2004JApA...25....1S} in the optical band.

Thermal infrared radiation in AGN is generally believed to be from the accretion disk and the dusty torus
outside of the dust sublimation radius having typical scales of around  
few tenths of parsec \citep{2018MNRAS.475.5330M,2014ApJ...788..159K}. Dust reverberation mapping observations show that the inner radius (R) of the torus ranges from 0.01 - 0.1 pc \citep{2006ApJ...639...46S}. Using this measured size of the torus, one can estimate the variability timescale ($\Delta t_{int}$) as 
\begin{equation}
R < c \times \Delta t_{int}
\end{equation}
Here, $\Delta t_{int} =  \Delta t_{obs}/(1+z)$ and $c$ is the speed of light. The FSRQs studied here have
redshifts between 0.19 to 3.10 and for those sources, using Equation 8 and
considering a $R$ of 0.1 pc, the observed timescale of variability ranges from
about 140 days (for $z$ = 0.19) to about 480 days (for $z$ = 3.10). In our analysis of intra-day
variability, the maximum time difference between the data points is 1.2 days. We have found
variations using such observations that span about 1.2 days and therefore our observed 
timescales of variability are much smaller than 1.2 days. This implies that the observed
mid infrared flux variations on intra-day timescales are not from the torus. Using the maximum time difference between
the data points (i.e. 1.2 days) as the observed variability timescale, from Equation 8,   
the mid 
infrared emission could have come from spatial scales ranging from 2 $\times$ 10$^{-4}$ (for a source at $z$ = 0.19)
to 8 $\times$ 10$^{-4}$ pc (for a source at $z$ = 3.10). This suggests that the observed mid infrared emission  
comes from a  very compact region corresponding to around few hundreds of Schwarzschild radii considering 
FSRQs to have black hole masses around 10$^8$ M$_{\odot}$. 
 The size of the emission region as deduced from from the observed intra-day variability timescale is much smaller than that of the torus and the accretion disk. This suggests that the observed radiation in W1 and W2 bands is dominated by the non-thermal 
emission from the jets of these sources. This  is not surprising, as these sources are  
$\gamma$-ray emitting sources and detected by {\it Fermi}. In this scenario, the observed 
mid infrared variability is well explained by jet based models 
\citep{1985ApJ...298..114M,2014ApJ...780...87M}. 

Statistical tests 
indicate FSRQs to be more variable than BL Lacs in both the W1 and W2 band on 
intra-day and long timescales. 
Analysis of the ensemble variability using structure function on long timescales shows differences in variability amplitude between
FSRQs and BL Lacs, which confirms the results obtained through amplitude of 
variability method.  Dividing the sample of sources based on the position of the synchrotron
peak in their broad band SED, structure function analysis and the analysis of variability amplitude too indicates that LSPs show mid infrared variability with the largest amplitude, followed by ISPs and HSPs. The increased variability of FSRQs relative to BL Lacs and LSPs relative to ISPs and HSPs could be due to a combination of the following two reasons.  Firstly, the observed W1 and W2 bands trace the peak of the synchrotron component in the case of FSRQs, while in the case of BL Lacs, it traces the rising part of the synchrotron component and thus the low energy electron population. Secondly, FSRQs are known to have
powerful jets compared to BL Lacs \citep{2018MNRAS.473.2639G}. 
As most of the FSRQs are LSP sources and BL Lacs
are HSP sources, LSPs show large amplitude variability compared to ISPs and HSPs.
Our results on the mid infrared amplitude of variability on long timescales is 
in agreement with what is found in the optical band. From an analysis of optical data in long term
\cite{2009ApJ...699.1732B} and \cite{2014MNRAS.439..690H} noticed FSRQs to be more variable than BL Lacs.
This too points to enhanced contribution of jet emission in the optical band of these sources.

\subsection{Colour variability}
Brightness variations in blazars are often accompanied by spectral variations
that manifest in colour-magnitude correlation plots or spectral index - magnitude
correlation plots. Analysis of spectral variations too are important as it can
provide additional clues to the origin of flux variations. Blazars have been
studied for spectral variations in the optical and near infrared bands 
\citep{2019MNRAS.484.5633G,2019ApJ...887..185S,2012ApJ...756...13B}, however, such studies in the
mid infrared band are very limited. For example, from an analysis of CTA 102 mid
infrared (W1 and W2) light curves during its optical 
outburst in 2016 \citep{2018A&A...617A..59K}, \cite{2018RNAAS...2..134J} noticed a BWB trend  within a day.
The data set analysed here is unique (larger number of sources and many epochs)
and it can provide insights on a statistical basis into the mid infrared colour variations in blazars.

The observed emission in W1 and W2 bands is a combination of thermal emission from the
accretion disk and torus and non-thermal synchrotron emission from the relativistic 
jet \citep{2016ApJ...817..119K}. For most of the $\gamma$-ray bright blazars, the bright state in the 
$\gamma$-ray band is accompanied by correlated increase in brightness in the optical 
as well as the infrared bands \citep{2012ApJ...756...13B}.
However, there are exceptions to this general
observations known in blazars. There are cases in blazars where the optical, near infrared  and $\gamma$-ray
emissions are not correlated \citep{2013ApJ...763L..11C,2019MNRAS.486.1781R,2019ApJ...880...32L}.
From a systematic study of long term optical flux variations in  $\gamma$-ray detected blazars
along with a control sample of gamma-ray undetected blazars, it has been found that
$\gamma$-ray detected blazars are  more variable than $\gamma$-ray undetected
blazars \citep{2014MNRAS.439..690H}.  Thus, it is clear that the observed 
flux variations (across the electromagnetic spectrum), both on intra-day 
timescales and long timescales in $\gamma$-ray bright blazars is due to the relativistic jets in them. 
The observed  BWB trend seen in
this study can be explained by 
 (i) localized temperature changes in the accretion disk with changes in the accretion rate \citep{2014ApJ...783..105R}, (ii) increased amplitude of variability at shorter wavelengths \citep{2009MNRAS.399.1357S} which in the one zone synchrotron emission model could happen due to the injection of fresh electrons that have an energy distribution harder than that of the earlier cooled electrons \citep{1998A&A...333..452K,2002PASA...19..138M} and (iii) changes in the Doppler factor in a convex spectrum \citep{2004A&A...421..103V} or variation in Doppler factor due to changes in the viewing angle of a curved and inhomogeneous jet \citep{2007A&A...470..857P}. Considering that the observed flux variations are intrinsic to the source, we rule out  geometric effects on the cause of colour variations and instead focus on the intrinsic causes for the observed colour variations. Among blazars, FSRQs have a stronger accretion disk and more powerful jets relative to BL Lacs that have weak accretion disk and less powerful jets \citep{2018MNRAS.473.2639G}. In both FSRQs and BL Lacs, predominantly BWB trend is observed and therefore, spectral variations with a BWB trend that could result from temperature changes in the accretion disk with changes in the accretion rate is disfavoured. Flux variability analysis discussed in Section 4.1 unambiguously point to the jet based origin of the observed infrared variability and therefore, the observed colour variations are related to complex processes intrinsic to the jets of both FSRQs and BL Lacs.

\section{Summary}

Using a large sample of 1035 blazars taken from the third catalog of AGN detected by
{\it Fermi}, and cross-matched with the WISE catalog, we studied the mid infrared variability 
properties of FSRQs and BL Lacs on both intra-day and long timescales. While blazars have
been studied for mid infrared variability on long timescales \citep{2018Ap&SS.363..167M}, the study presented here
is the first on the mid infrared intra-day variability characteristics of different categories of
blazars using an extensive data set taken from WISE observations.  
We quantified variability by calculating the amplitude of variability, $\sigma_m$.  
In addition to flux variability, we also studied the infrared 
colour variations in our sample of sources. The major findings of this present study are summarized below

\begin{enumerate}
\item All sources in our sample, except three showed flux variations both on intra-day and long timescales.
\item On intra-day timescales, we found FSRQs to show larger amplitude flux 
variations in the mid infrared W1 and W2 bands relative to BL Lac objects.  
When the sample is divided into different sub-classes based on the 
position of the synchrotron peak in their broad band SEDs, LSPs showed the largest
amplitude of variability while HSPs and ISPs showed similar variability amplitudes. 
\item On long timescales, FSRQs showed large amplitude flux variations compared to BL Lacs in
both W1 and W2 bands. However, there is no difference in variability amplitudes between
W1 and W2 bands in FSRQs and BL Lacs. Among the 
various sub-classes of blazars, in W1 and W2 bands, LSPs showed the largest 
amplitude of flux variability and HSPs showed
the lowest amplitude of flux variations, while ISP sources sources showed flux variations with 
the amplitude of variability intermediate between LSPs and HSPs. 
\item Most of FSRQs and BL Lacs showed a bluer when brighter trend, while only a small fraction of them showed a 
redder when brighter
behaviour.
\item No correlation was found between the mid infrared amplitude of variability and black hole mass in both FSRQs
and BL Lacs.
\item From the analysis of intra-day light curves, we found BL Lacs to show an 
increased duty cycle of variability than FSRQs in both W1 and W2 bands.

\end{enumerate}

\section*{Acknowledgements}
 We thank the anonymous referee for his/her critical review of our manuscript that
helped to improve the presentation significantly. This publication makes use of 
data products from the Wide-field Infrared Survey 
Explorer, which is a joint project of the University of California, Los Angeles,
and the Jet Propulsion Laboratory/California Institute of Technology, funded by 
the National Aeronautics and Space Administration.
%





\bibliographystyle{mnras}
\bibliography{ref}

\begin{thebibliography}{}
\makeatletter
\relax
\def\mn@urlcharsother{\let\do\@makeother \do\$\do\&\do\#\do\^\do\_\do\%\do\~}
\def\mn@doi{\begingroup\mn@urlcharsother \@ifnextchar [ {\mn@doi@}
  {\mn@doi@[]}}
\def\mn@doi@[#1]#2{\def\@tempa{#1}\ifx\@tempa\@empty \href
  {http://dx.doi.org/#2} {doi:#2}\else \href {http://dx.doi.org/#2} {#1}\fi
  \endgroup}
\def\mn@eprint#1#2{\mn@eprint@#1:#2::\@nil}
\def\mn@eprint@arXiv#1{\href {http://arxiv.org/abs/#1} {{\tt arXiv:#1}}}
\def\mn@eprint@dblp#1{\href {http://dblp.uni-trier.de/rec/bibtex/#1.xml}
  {dblp:#1}}
\def\mn@eprint@#1:#2:#3:#4\@nil{\def\@tempa {#1}\def\@tempb {#2}\def\@tempc
  {#3}\ifx \@tempc \@empty \let \@tempc \@tempb \let \@tempb \@tempa \fi \ifx
  \@tempb \@empty \def\@tempb {arXiv}\fi \@ifundefined
  {mn@eprint@\@tempb}{\@tempb:\@tempc}{\expandafter \expandafter \csname
  mn@eprint@\@tempb\endcsname \expandafter{\@tempc}}}

\bibitem[\protect\citeauthoryear{{Abdo} et~al.,}{{Abdo}
  et~al.}{2010}]{2010ApJ...716...30A}
{Abdo} A.~A.,  et~al., 2010, \mn@doi [\apj] {10.1088/0004-637X/716/1/30}, \href
  {https://ui.adsabs.harvard.edu/abs/2010ApJ...716...30A} {716, 30}

\bibitem[\protect\citeauthoryear{{Ackermann} et~al.,}{{Ackermann}
  et~al.}{2015}]{2015ApJ...810...14A}
{Ackermann} M.,  et~al., 2015, \mn@doi [\apj] {10.1088/0004-637X/810/1/14},
  \href {https://ui.adsabs.harvard.edu/abs/2015ApJ...810...14A} {810, 14}

\bibitem[\protect\citeauthoryear{{Andruchow}, {Romero}  \&
  {Cellone}}{{Andruchow} et~al.}{2005}]{2005A&A...442...97A}
{Andruchow} I.,  {Romero} G.~E.,   {Cellone} S.~A.,  2005, \mn@doi [\aap]
  {10.1051/0004-6361:20053325}, \href
  {https://ui.adsabs.harvard.edu/abs/2005A&A...442...97A} {442, 97}

\bibitem[\protect\citeauthoryear{{Atwood} et~al.,}{{Atwood}
  et~al.}{2009}]{2009ApJ...697.1071A}
{Atwood} W.~B.,  et~al., 2009, \mn@doi [\apj] {10.1088/0004-637X/697/2/1071},
  \href {https://ui.adsabs.harvard.edu/abs/2009ApJ...697.1071A} {697, 1071}

\bibitem[\protect\citeauthoryear{{Bauer}, {Baltay}, {Coppi}, {Ellman}, {Jerke},
  {Rabinowitz}  \& {Scalzo}}{{Bauer} et~al.}{2009}]{2009ApJ...699.1732B}
{Bauer} A.,  {Baltay} C.,  {Coppi} P.,  {Ellman} N.,  {Jerke} J.,  {Rabinowitz}
  D.,   {Scalzo} R.,  2009, \mn@doi [\apj] {10.1088/0004-637X/699/2/1732},
  \href {https://ui.adsabs.harvard.edu/abs/2009ApJ...699.1732B} {699, 1732}

\bibitem[\protect\citeauthoryear{{Begelman} et~al.,}{{Begelman}
  et~al.}{1987}]{1987ApJ...322..650B}
{Begelman} M.~C.,  et~al., 1987, \mn@doi [\apj] {10.1086/165760}, \href
  {https://ui.adsabs.harvard.edu/abs/1987ApJ...322..650B} {322, 650}

\bibitem[\protect\citeauthoryear{{Bonning} et~al.,}{{Bonning}
  et~al.}{2012}]{2012ApJ...756...13B}
{Bonning} E.,  et~al., 2012, \mn@doi [\apj] {10.1088/0004-637X/756/1/13}, \href
  {https://ui.adsabs.harvard.edu/abs/2012ApJ...756...13B} {756, 13}

\bibitem[\protect\citeauthoryear{{B{\"o}ttcher}, {Reimer}, {Sweeney}  \&
  {Prakash}}{{B{\"o}ttcher} et~al.}{2013}]{2013ApJ...768...54B}
{B{\"o}ttcher} M.,  {Reimer} A.,  {Sweeney} K.,   {Prakash} A.,  2013, \mn@doi
  [\apj] {10.1088/0004-637X/768/1/54}, \href
  {https://ui.adsabs.harvard.edu/abs/2013ApJ...768...54B} {768, 54}

\bibitem[\protect\citeauthoryear{{Carnerero} et~al.,}{{Carnerero}
  et~al.}{2015}]{2015MNRAS.450.2677C}
{Carnerero} M.~I.,  et~al., 2015, \mn@doi [\mnras] {10.1093/mnras/stv823},
  \href {https://ui.adsabs.harvard.edu/abs/2015MNRAS.450.2677C} {450, 2677}

\bibitem[\protect\citeauthoryear{{Chatterjee} et~al.,}{{Chatterjee}
  et~al.}{2013}]{2013ApJ...763L..11C}
{Chatterjee} R.,  et~al., 2013, \mn@doi [\apjl] {10.1088/2041-8205/763/1/L11},
  \href {https://ui.adsabs.harvard.edu/abs/2013ApJ...763L..11C} {763, L11}

\bibitem[\protect\citeauthoryear{{Dermer}, {Schlickeiser}  \&
  {Mastichiadis}}{{Dermer} et~al.}{1992}]{1992A&A...256L..27D}
{Dermer} C.~D.,  {Schlickeiser} R.,   {Mastichiadis} A.,  1992, \aap, \href
  {https://ui.adsabs.harvard.edu/abs/1992A&A...256L..27D} {256, L27}

\bibitem[\protect\citeauthoryear{{Edelson}}{{Edelson}}{1992}]{1992ApJ...401..5%
16E}
{Edelson} R.,  1992, \mn@doi [\apj] {10.1086/172083}, \href
  {https://ui.adsabs.harvard.edu/abs/1992ApJ...401..516E} {401, 516}

\bibitem[\protect\citeauthoryear{{Edelson} et~al.,}{{Edelson}
  et~al.}{1991}]{1991ApJ...372L...9E}
{Edelson} R.~A.,  et~al., 1991, \mn@doi [\apjl] {10.1086/186011}, \href
  {https://ui.adsabs.harvard.edu/abs/1991ApJ...372L...9E} {372, L9}

\bibitem[\protect\citeauthoryear{{Emmanoulopoulos}, {McHardy}  \&
  {Uttley}}{{Emmanoulopoulos} et~al.}{2010}]{2010MNRAS.404..931E}
{Emmanoulopoulos} D.,  {McHardy} I.~M.,   {Uttley} P.,  2010, \mn@doi [\mnras]
  {10.1111/j.1365-2966.2010.16328.x}, \href
  {https://ui.adsabs.harvard.edu/abs/2010MNRAS.404..931E} {404, 931}

\bibitem[\protect\citeauthoryear{{Fanaroff} \& {Riley}}{{Fanaroff} \&
  {Riley}}{1974}]{1974MNRAS.167P..31F}
{Fanaroff} B.~L.,  {Riley} J.~M.,  1974, \mn@doi [\mnras]
  {10.1093/mnras/167.1.31P}, \href
  {https://ui.adsabs.harvard.edu/abs/1974MNRAS.167P..31F} {167, 31P}

\bibitem[\protect\citeauthoryear{{Gab{\'a}nyi}, {Mo{\'o}r}  \&
  {Frey}}{{Gab{\'a}nyi} et~al.}{2018}]{2018RNAAS...2..130G}
{Gab{\'a}nyi} K.~{\'E}.,  {Mo{\'o}r} A.,   {Frey} S.,  2018, \mn@doi [Research
  Notes of the American Astronomical Society] {10.3847/2515-5172/aad49f}, \href
  {https://ui.adsabs.harvard.edu/abs/2018RNAAS...2..130G} {2, 130}

\bibitem[\protect\citeauthoryear{{Gardner} \& {Done}}{{Gardner} \&
  {Done}}{2018}]{2018MNRAS.473.2639G}
{Gardner} E.,  {Done} C.,  2018, \mn@doi [\mnras] {10.1093/mnras/stx2516},
  \href {https://ui.adsabs.harvard.edu/abs/2018MNRAS.473.2639G} {473, 2639}

\bibitem[\protect\citeauthoryear{{Gaur} et~al.,}{{Gaur}
  et~al.}{2012}]{2012MNRAS.425.3002G}
{Gaur} H.,  et~al., 2012, \mn@doi [\mnras] {10.1111/j.1365-2966.2012.21583.x},
  \href {https://ui.adsabs.harvard.edu/abs/2012MNRAS.425.3002G} {425, 3002}

\bibitem[\protect\citeauthoryear{{Gaur} et~al.,}{{Gaur}
  et~al.}{2019}]{2019MNRAS.484.5633G}
{Gaur} H.,  et~al., 2019, \mn@doi [\mnras] {10.1093/mnras/stz322}, \href
  {https://ui.adsabs.harvard.edu/abs/2019MNRAS.484.5633G} {484, 5633}

\bibitem[\protect\citeauthoryear{{Ghisellini} \& {Maraschi}}{{Ghisellini} \&
  {Maraschi}}{1989}]{1989ApJ...340..181G}
{Ghisellini} G.,  {Maraschi} L.,  1989, \mn@doi [\apj] {10.1086/167383}, \href
  {https://ui.adsabs.harvard.edu/abs/1989ApJ...340..181G} {340, 181}

\bibitem[\protect\citeauthoryear{{Ghisellini}, {Tavecchio}, {Foschini}  \&
  {Ghirland a}}{{Ghisellini} et~al.}{2011}]{2011MNRAS.414.2674G}
{Ghisellini} G.,  {Tavecchio} F.,  {Foschini} L.,   {Ghirland a} G.,  2011,
  \mn@doi [\mnras] {10.1111/j.1365-2966.2011.18578.x}, \href
  {https://ui.adsabs.harvard.edu/abs/2011MNRAS.414.2674G} {414, 2674}

\bibitem[\protect\citeauthoryear{{Gu} \& {Ai}}{{Gu} \&
  {Ai}}{2011}]{2011A&A...528A..95G}
{Gu} M.~F.,  {Ai} Y.~L.,  2011, \mn@doi [\aap] {10.1051/0004-6361/201016280},
  \href {https://ui.adsabs.harvard.edu/abs/2011A&A...528A..95G} {528, A95}

\bibitem[\protect\citeauthoryear{{Gu}, {Lee}, {Pak}, {Yim}  \& {Fletcher}}{{Gu}
  et~al.}{2006}]{2006A&A...450...39G}
{Gu} M.~F.,  {Lee} C.~U.,  {Pak} S.,  {Yim} H.~S.,   {Fletcher} A.~B.,  2006,
  \mn@doi [\aap] {10.1051/0004-6361:20054271}, \href
  {https://ui.adsabs.harvard.edu/abs/2006A&A...450...39G} {450, 39}

\bibitem[\protect\citeauthoryear{{Hovatta} et~al.,}{{Hovatta}
  et~al.}{2014}]{2014MNRAS.439..690H}
{Hovatta} T.,  et~al., 2014, \mn@doi [\mnras] {10.1093/mnras/stt2494}, \href
  {https://ui.adsabs.harvard.edu/abs/2014MNRAS.439..690H} {439, 690}

\bibitem[\protect\citeauthoryear{{Jarrett} et~al.,}{{Jarrett}
  et~al.}{2011}]{2011ApJ...735..112J}
{Jarrett} T.~H.,  et~al., 2011, \mn@doi [\apj] {10.1088/0004-637X/735/2/112},
  \href {https://ui.adsabs.harvard.edu/abs/2011ApJ...735..112J} {735, 112}

\bibitem[\protect\citeauthoryear{{Jiang}}{{Jiang}}{2018}]{2018RNAAS...2..134J}
{Jiang} N.,  2018, \mn@doi [Research Notes of the American Astronomical
  Society] {10.3847/2515-5172/aad693}, \href
  {https://ui.adsabs.harvard.edu/abs/2018RNAAS...2..134J} {2, 134}

\bibitem[\protect\citeauthoryear{{Jorstad} et~al.,}{{Jorstad}
  et~al.}{2005}]{2005AJ....130.1418J}
{Jorstad} S.~G.,  et~al., 2005, \mn@doi [\aj] {10.1086/444593}, \href
  {https://ui.adsabs.harvard.edu/abs/2005AJ....130.1418J} {130, 1418}

\bibitem[\protect\citeauthoryear{{Kaur} \& {Baliyan}}{{Kaur} \&
  {Baliyan}}{2018}]{2018A&A...617A..59K}
{Kaur} N.,  {Baliyan} K.~S.,  2018, \mn@doi [\aap]
  {10.1051/0004-6361/201731953}, \href
  {https://ui.adsabs.harvard.edu/abs/2018A&A...617A..59K} {617, A59}

\bibitem[\protect\citeauthoryear{{Kelly}, {Bechtold}  \&
  {Siemiginowska}}{{Kelly} et~al.}{2009}]{2009ApJ...698..895K}
{Kelly} B.~C.,  {Bechtold} J.,   {Siemiginowska} A.,  2009, \mn@doi [\apj]
  {10.1088/0004-637X/698/1/895}, \href
  {https://ui.adsabs.harvard.edu/abs/2009ApJ...698..895K} {698, 895}

\bibitem[\protect\citeauthoryear{{Kirk}, {Rieger}  \& {Mastichiadis}}{{Kirk}
  et~al.}{1998}]{1998A&A...333..452K}
{Kirk} J.~G.,  {Rieger} F.~M.,   {Mastichiadis} A.,  1998, \aap, \href
  {https://ui.adsabs.harvard.edu/abs/1998A&A...333..452K} {333, 452}

\bibitem[\protect\citeauthoryear{{Konigl}}{{Konigl}}{1981}]{1981ApJ...243..700%
K}
{Konigl} A.,  1981, \mn@doi [\apj] {10.1086/158638}, \href
  {https://ui.adsabs.harvard.edu/abs/1981ApJ...243..700K} {243, 700}

\bibitem[\protect\citeauthoryear{{Koshida} et~al.,}{{Koshida}
  et~al.}{2014}]{2014ApJ...788..159K}
{Koshida} S.,  et~al., 2014, \mn@doi [\apj] {10.1088/0004-637X/788/2/159},
  \href {https://ui.adsabs.harvard.edu/abs/2014ApJ...788..159K} {788, 159}

\bibitem[\protect\citeauthoryear{{Koz{\l}owski}}{{Koz{\l}owski}}{2016}]{2016Ap%
J...826..118K}
{Koz{\l}owski} S.,  2016, \mn@doi [\apj] {10.3847/0004-637X/826/2/118}, \href
  {https://ui.adsabs.harvard.edu/abs/2016ApJ...826..118K} {826, 118}

\bibitem[\protect\citeauthoryear{{Koz{\l}owski}, {Kochanek}, {Ashby}, {Assef},
  {Brodwin}, {Eisenhardt}, {Jannuzi}  \& {Stern}}{{Koz{\l}owski}
  et~al.}{2016}]{2016ApJ...817..119K}
{Koz{\l}owski} S.,  {Kochanek} C.~S.,  {Ashby} M. L.~N.,  {Assef} R.~J.,
  {Brodwin} M.,  {Eisenhardt} P.~R.,  {Jannuzi} B.~T.,   {Stern} D.,  2016,
  \mn@doi [\apj] {10.3847/0004-637X/817/2/119}, \href
  {https://ui.adsabs.harvard.edu/abs/2016ApJ...817..119K} {817, 119}

\bibitem[\protect\citeauthoryear{{Liodakis}, {Romani}, {Filippenko},
  {Kiehlmann}, {Max-Moerbeck}, {Readhead}  \& {Zheng}}{{Liodakis}
  et~al.}{2018}]{2018MNRAS.480.5517L}
{Liodakis} I.,  {Romani} R.~W.,  {Filippenko} A.~V.,  {Kiehlmann} S.,
  {Max-Moerbeck} W.,  {Readhead} A.~C.~S.,   {Zheng} W.,  2018, \mn@doi
  [\mnras] {10.1093/mnras/sty2264}, \href
  {https://ui.adsabs.harvard.edu/abs/2018MNRAS.480.5517L} {480, 5517}

\bibitem[\protect\citeauthoryear{{Liodakis}, {Romani}, {Filippenko}, {Kocevski}
   \& {Zheng}}{{Liodakis} et~al.}{2019}]{2019ApJ...880...32L}
{Liodakis} I.,  {Romani} R.~W.,  {Filippenko} A.~V.,  {Kocevski} D.,   {Zheng}
  W.,  2019, \mn@doi [\apj] {10.3847/1538-4357/ab26b7}, \href
  {https://ui.adsabs.harvard.edu/abs/2019ApJ...880...32L} {880, 32}

\bibitem[\protect\citeauthoryear{{Lu} et~al.,}{{Lu}
  et~al.}{2019}]{2019ApJ...877...23L}
{Lu} K.-X.,  et~al., 2019, \mn@doi [\apj] {10.3847/1538-4357/ab16e8}, \href
  {https://ui.adsabs.harvard.edu/abs/2019ApJ...877...23L} {877, 23}

\bibitem[\protect\citeauthoryear{{Lynden-Bell}}{{Lynden-Bell}}{1969}]{1969Natu%
r.223..690L}
{Lynden-Bell} D.,  1969, \mn@doi [\nat] {10.1038/223690a0}, \href
  {https://ui.adsabs.harvard.edu/abs/1969Natur.223..690L} {223, 690}

\bibitem[\protect\citeauthoryear{{Mainzer} et~al.,}{{Mainzer}
  et~al.}{2011}]{2011ApJ...731...53M}
{Mainzer} A.,  et~al., 2011, \mn@doi [\apj] {10.1088/0004-637X/731/1/53}, \href
  {https://ui.adsabs.harvard.edu/abs/2011ApJ...731...53M} {731, 53}

\bibitem[\protect\citeauthoryear{{Mandal} et~al.,}{{Mandal}
  et~al.}{2018}]{2018MNRAS.475.5330M}
{Mandal} A.~K.,  et~al., 2018, \mn@doi [\mnras] {10.1093/mnras/sty200}, \href
  {https://ui.adsabs.harvard.edu/abs/2018MNRAS.475.5330M} {475, 5330}

\bibitem[\protect\citeauthoryear{{Mao}, {Zhang}  \& {Yi}}{{Mao}
  et~al.}{2018}]{2018Ap&SS.363..167M}
{Mao} L.,  {Zhang} X.,   {Yi} T.,  2018, \mn@doi [\apss]
  {10.1007/s10509-018-3388-9}, \href
  {https://ui.adsabs.harvard.edu/abs/2018Ap&SS.363..167M} {363, 167}

\bibitem[\protect\citeauthoryear{{Marscher}}{{Marscher}}{2014}]{2014ApJ...780.%
..87M}
{Marscher} A.~P.,  2014, \mn@doi [\apj] {10.1088/0004-637X/780/1/87}, \href
  {https://ui.adsabs.harvard.edu/abs/2014ApJ...780...87M} {780, 87}

\bibitem[\protect\citeauthoryear{{Marscher} \& {Gear}}{{Marscher} \&
  {Gear}}{1985}]{1985ApJ...298..114M}
{Marscher} A.~P.,  {Gear} W.~K.,  1985, \mn@doi [\apj] {10.1086/163592}, \href
  {https://ui.adsabs.harvard.edu/abs/1985ApJ...298..114M} {298, 114}

\bibitem[\protect\citeauthoryear{{Massaro}, {Nesci}, {Maesano}, {Montagni}  \&
  {D'Alessio}}{{Massaro} et~al.}{1998}]{1998MNRAS.299...47M}
{Massaro} E.,  {Nesci} R.,  {Maesano} M.,  {Montagni} F.,   {D'Alessio} F.,
  1998, \mn@doi [\mnras] {10.1046/j.1365-8711.1998.01696.x}, \href
  {https://ui.adsabs.harvard.edu/abs/1998MNRAS.299...47M} {299, 47}

\bibitem[\protect\citeauthoryear{{Mastichiadis} \& {Kirk}}{{Mastichiadis} \&
  {Kirk}}{2002}]{2002PASA...19..138M}
{Mastichiadis} A.,  {Kirk} J.~G.,  2002, \mn@doi [\pasa] {10.1071/AS01108},
  \href {https://ui.adsabs.harvard.edu/abs/2002PASA...19..138M} {19, 138}

\bibitem[\protect\citeauthoryear{{Melia} \& {Konigl}}{{Melia} \&
  {Konigl}}{1989}]{1989ApJ...340..162M}
{Melia} F.,  {Konigl} A.,  1989, \mn@doi [\apj] {10.1086/167382}, \href
  {https://ui.adsabs.harvard.edu/abs/1989ApJ...340..162M} {340, 162}

\bibitem[\protect\citeauthoryear{{Paliya}, {B{\"o}ttcher}, {Diltz}, {Stalin},
  {Sahayanathan}  \& {Ravikumar}}{{Paliya} et~al.}{2015}]{2015ApJ...811..143P}
{Paliya} V.~S.,  {B{\"o}ttcher} M.,  {Diltz} C.,  {Stalin} C.~S.,
  {Sahayanathan} S.,   {Ravikumar} C.~D.,  2015, \mn@doi [\apj]
  {10.1088/0004-637X/811/2/143}, \href
  {https://ui.adsabs.harvard.edu/abs/2015ApJ...811..143P} {811, 143}

\bibitem[\protect\citeauthoryear{{Paliya}, {Diltz}, {B{\"o}ttcher}, {Stalin}
  \& {Buckley}}{{Paliya} et~al.}{2016}]{2016ApJ...817...61P}
{Paliya} V.~S.,  {Diltz} C.,  {B{\"o}ttcher} M.,  {Stalin} C.~S.,   {Buckley}
  D.,  2016, \mn@doi [\apj] {10.3847/0004-637X/817/1/61}, \href
  {https://ui.adsabs.harvard.edu/abs/2016ApJ...817...61P} {817, 61}

\bibitem[\protect\citeauthoryear{{Papadakis}, {Villata}  \&
  {Raiteri}}{{Papadakis} et~al.}{2007}]{2007A&A...470..857P}
{Papadakis} I.~E.,  {Villata} M.,   {Raiteri} C.~M.,  2007, \mn@doi [\aap]
  {10.1051/0004-6361:20077516}, \href
  {https://ui.adsabs.harvard.edu/abs/2007A&A...470..857P} {470, 857}

\bibitem[\protect\citeauthoryear{{Rajput}, {Stalin}, {Sahayanathan}, {Rakshit}
  \& {Mandal}}{{Rajput} et~al.}{2019}]{2019MNRAS.486.1781R}
{Rajput} B.,  {Stalin} C.~S.,  {Sahayanathan} S.,  {Rakshit} S.,   {Mandal}
  A.~K.,  2019, \mn@doi [\mnras] {10.1093/mnras/stz941}, \href
  {https://ui.adsabs.harvard.edu/abs/2019MNRAS.486.1781R} {486, 1781}

\bibitem[\protect\citeauthoryear{{Rakshit}, {Stalin}, {Muneer}, {Neha}  \&
  {Paliya}}{{Rakshit} et~al.}{2017}]{2017ApJ...835..275R}
{Rakshit} S.,  {Stalin} C.~S.,  {Muneer} S.,  {Neha} S.,   {Paliya} V.~S.,
  2017, \mn@doi [\apj] {10.3847/1538-4357/835/2/275}, \href
  {https://ui.adsabs.harvard.edu/abs/2017ApJ...835..275R} {835, 275}

\bibitem[\protect\citeauthoryear{{Rakshit}, {Johnson}, {Stalin}, {Gand hi}  \&
  {Hoenig}}{{Rakshit} et~al.}{2019}]{2019MNRAS.483.2362R}
{Rakshit} S.,  {Johnson} A.,  {Stalin} C.~S.,  {Gand hi} P.,   {Hoenig} S.,
  2019, \mn@doi [\mnras] {10.1093/mnras/sty3261}, \href
  {https://ui.adsabs.harvard.edu/abs/2019MNRAS.483.2362R} {483, 2362}

\bibitem[\protect\citeauthoryear{{Rani}, {Stalin}  \& {Rakshit}}{{Rani}
  et~al.}{2017}]{2017MNRAS.466.3309R}
{Rani} P.,  {Stalin} C.~S.,   {Rakshit} S.,  2017, \mn@doi [\mnras]
  {10.1093/mnras/stw3228}, \href
  {https://ui.adsabs.harvard.edu/abs/2017MNRAS.466.3309R} {466, 3309}

\bibitem[\protect\citeauthoryear{{Rees}}{{Rees}}{1984}]{1984ARA&A..22..471R}
{Rees} M.~J.,  1984, \mn@doi [\araa] {10.1146/annurev.aa.22.090184.002351},
  \href {https://ui.adsabs.harvard.edu/abs/1984ARA&A..22..471R} {22, 471}

\bibitem[\protect\citeauthoryear{{Romero}, {Cellone}  \& {Combi}}{{Romero}
  et~al.}{1999}]{1999A&AS..135..477R}
{Romero} G.~E.,  {Cellone} S.~A.,   {Combi} J.~A.,  1999, \mn@doi [\aaps]
  {10.1051/aas:1999184}, \href
  {https://ui.adsabs.harvard.edu/abs/1999A&AS..135..477R} {135, 477}

\bibitem[\protect\citeauthoryear{{Ruan}, {Anderson}, {Dexter}  \&
  {Agol}}{{Ruan} et~al.}{2014}]{2014ApJ...783..105R}
{Ruan} J.~J.,  {Anderson} S.~F.,  {Dexter} J.,   {Agol} E.,  2014, \mn@doi
  [\apj] {10.1088/0004-637X/783/2/105}, \href
  {https://ui.adsabs.harvard.edu/abs/2014ApJ...783..105R} {783, 105}

\bibitem[\protect\citeauthoryear{{Sarkar} et~al.,}{{Sarkar}
  et~al.}{2019}]{2019ApJ...887..185S}
{Sarkar} A.,  et~al., 2019, \mn@doi [\apj] {10.3847/1538-4357/ab5281}, \href
  {https://ui.adsabs.harvard.edu/abs/2019ApJ...887..185S} {887, 185}

\bibitem[\protect\citeauthoryear{{Sesar} et~al.,}{{Sesar}
  et~al.}{2007}]{2007AJ....134.2236S}
{Sesar} B.,  et~al., 2007, \mn@doi [\aj] {10.1086/521819}, \href
  {https://ui.adsabs.harvard.edu/abs/2007AJ....134.2236S} {134, 2236}

\bibitem[\protect\citeauthoryear{{Simm}, {Salvato}, {Saglia}, {Ponti},
  {Lanzuisi}, {Trakhtenbrot}, {Nandra}  \& {Bender}}{{Simm}
  et~al.}{2016}]{2016A&A...585A.129S}
{Simm} T.,  {Salvato} M.,  {Saglia} R.,  {Ponti} G.,  {Lanzuisi} G.,
  {Trakhtenbrot} B.,  {Nandra} K.,   {Bender} R.,  2016, \mn@doi [\aap]
  {10.1051/0004-6361/201527353}, \href
  {https://ui.adsabs.harvard.edu/abs/2016A&A...585A.129S} {585, A129}

\bibitem[\protect\citeauthoryear{{Simonetti}, {Cordes}  \&
  {Heeschen}}{{Simonetti} et~al.}{1985}]{1985ApJ...296...46S}
{Simonetti} J.~H.,  {Cordes} J.~M.,   {Heeschen} D.~S.,  1985, \mn@doi [\apj]
  {10.1086/163418}, \href
  {https://ui.adsabs.harvard.edu/abs/1985ApJ...296...46S} {296, 46}

\bibitem[\protect\citeauthoryear{{Stalin}, {Gopal-Krishna}, {Sagar}  \&
  {Wiita}}{{Stalin} et~al.}{2004a}]{2004JApA...25....1S}
{Stalin} C.~S.,  {Gopal-Krishna} {Sagar} R.,   {Wiita} P.~J.,  2004a, \mn@doi
  [Journal of Astrophysics and Astronomy] {10.1007/BF02702287}, \href
  {https://ui.adsabs.harvard.edu/abs/2004JApA...25....1S} {25, 1}

\bibitem[\protect\citeauthoryear{{Stalin}, {Gopal-Krishna}, {Sagar}  \&
  {Wiita}}{{Stalin} et~al.}{2004b}]{2004MNRAS.350..175S}
{Stalin} C.~S.,  {Gopal-Krishna} {Sagar} R.,   {Wiita} P.~J.,  2004b, \mn@doi
  [\mnras] {10.1111/j.1365-2966.2004.07631.x}, \href
  {https://ui.adsabs.harvard.edu/abs/2004MNRAS.350..175S} {350, 175}

\bibitem[\protect\citeauthoryear{{Stalin}, {Gopal-Krishna}, {Sagar}, {Wiita},
  {Mohan}  \& {Pandey}}{{Stalin} et~al.}{2006}]{2006MNRAS.366.1337S}
{Stalin} C.~S.,  {Gopal-Krishna} {Sagar} R.,  {Wiita} P.~J.,  {Mohan} V.,
  {Pandey} A.~K.,  2006, \mn@doi [\mnras] {10.1111/j.1365-2966.2005.09939.x},
  \href {https://ui.adsabs.harvard.edu/abs/2006MNRAS.366.1337S} {366, 1337}

\bibitem[\protect\citeauthoryear{{Stalin} et~al.,}{{Stalin}
  et~al.}{2009}]{2009MNRAS.399.1357S}
{Stalin} C.~S.,  et~al., 2009, \mn@doi [\mnras]
  {10.1111/j.1365-2966.2009.15354.x}, \href
  {https://ui.adsabs.harvard.edu/abs/2009MNRAS.399.1357S} {399, 1357}

\bibitem[\protect\citeauthoryear{{Stickel}, {Padovani}, {Urry}, {Fried}  \&
  {Kuehr}}{{Stickel} et~al.}{1991}]{1991ApJ...374..431S}
{Stickel} M.,  {Padovani} P.,  {Urry} C.~M.,  {Fried} J.~W.,   {Kuehr} H.,
  1991, \mn@doi [\apj] {10.1086/170133}, \href
  {https://ui.adsabs.harvard.edu/abs/1991ApJ...374..431S} {374, 431}

\bibitem[\protect\citeauthoryear{{Stocke}, {Morris}, {Gioia}, {Maccacaro},
  {Schild}, {Wolter}, {Fleming}  \& {Henry}}{{Stocke}
  et~al.}{1991}]{1991ApJS...76..813S}
{Stocke} J.~T.,  {Morris} S.~L.,  {Gioia} I.~M.,  {Maccacaro} T.,  {Schild} R.,
   {Wolter} A.,  {Fleming} T.~A.,   {Henry} J.~P.,  1991, \mn@doi [\apjs]
  {10.1086/191582}, \href
  {https://ui.adsabs.harvard.edu/abs/1991ApJS...76..813S} {76, 813}

\bibitem[\protect\citeauthoryear{{Suganuma} et~al.,}{{Suganuma}
  et~al.}{2006}]{2006ApJ...639...46S}
{Suganuma} M.,  et~al., 2006, \mn@doi [\apj] {10.1086/499326}, \href
  {https://ui.adsabs.harvard.edu/abs/2006ApJ...639...46S} {639, 46}

\bibitem[\protect\citeauthoryear{{Sukanya}, {Stalin}, {Jeyakumar}, {Praveen},
  {Dhani}  \& {Damle}}{{Sukanya} et~al.}{2016}]{2016RAA....16...27S}
{Sukanya} N.,  {Stalin} C.~S.,  {Jeyakumar} S.,  {Praveen} D.,  {Dhani} A.,
  {Damle} R.,  2016, \mn@doi [Research in Astronomy and Astrophysics]
  {10.1088/1674-4527/16/2/027}, \href
  {https://ui.adsabs.harvard.edu/abs/2016RAA....16...27S} {16, 27}

\bibitem[\protect\citeauthoryear{{The Fermi-LAT collaboration}}{{The Fermi-LAT
  collaboration}}{2019}]{2019arXiv190210045T}
{The Fermi-LAT collaboration} 2019, arXiv e-prints, \href
  {https://ui.adsabs.harvard.edu/abs/2019arXiv190210045T} {p. arXiv:1902.10045}

\bibitem[\protect\citeauthoryear{{Vagnetti}, {Trevese}  \& {Nesci}}{{Vagnetti}
  et~al.}{2003}]{2003ApJ...590..123V}
{Vagnetti} F.,  {Trevese} D.,   {Nesci} R.,  2003, \mn@doi [\apj]
  {10.1086/374889}, \href
  {https://ui.adsabs.harvard.edu/abs/2003ApJ...590..123V} {590, 123}

\bibitem[\protect\citeauthoryear{{Vanden Berk} et~al.,}{{Vanden Berk}
  et~al.}{2004}]{2004ApJ...601..692V}
{Vanden Berk} D.~E.,  et~al., 2004, \mn@doi [\apj] {10.1086/380563}, \href
  {https://ui.adsabs.harvard.edu/abs/2004ApJ...601..692V} {601, 692}

\bibitem[\protect\citeauthoryear{{Villata} et~al.,}{{Villata}
  et~al.}{2002}]{2002A&A...390..407V}
{Villata} M.,  et~al., 2002, \mn@doi [\aap] {10.1051/0004-6361:20020662}, \href
  {https://ui.adsabs.harvard.edu/abs/2002A&A...390..407V} {390, 407}

\bibitem[\protect\citeauthoryear{{Villata} et~al.,}{{Villata}
  et~al.}{2004}]{2004A&A...421..103V}
{Villata} M.,  et~al., 2004, \mn@doi [\aap] {10.1051/0004-6361:20035895}, \href
  {https://ui.adsabs.harvard.edu/abs/2004A&A...421..103V} {421, 103}

\bibitem[\protect\citeauthoryear{{Welsh}, {Wheatley}  \& {Neil}}{{Welsh}
  et~al.}{2011}]{2011A&A...527A..15W}
{Welsh} B.~Y.,  {Wheatley} J.~M.,   {Neil} J.~D.,  2011, \mn@doi [\aap]
  {10.1051/0004-6361/201015865}, \href
  {https://ui.adsabs.harvard.edu/abs/2011A&A...527A..15W} {527, A15}

\bibitem[\protect\citeauthoryear{{Wold}, {Brotherton}  \& {Shang}}{{Wold}
  et~al.}{2007}]{2007MNRAS.375..989W}
{Wold} M.,  {Brotherton} M.~S.,   {Shang} Z.,  2007, \mn@doi [\mnras]
  {10.1111/j.1365-2966.2006.11364.x}, \href
  {https://ui.adsabs.harvard.edu/abs/2007MNRAS.375..989W} {375, 989}

\bibitem[\protect\citeauthoryear{{Wright} et~al.,}{{Wright}
  et~al.}{2010}]{2010AJ....140.1868W}
{Wright} E.~L.,  et~al., 2010, \mn@doi [\aj] {10.1088/0004-6256/140/6/1868},
  \href {https://ui.adsabs.harvard.edu/abs/2010AJ....140.1868W} {140, 1868}

\bibitem[\protect\citeauthoryear{{Wu}, {Zhou}, {Ma}  \& {Jiang}}{{Wu}
  et~al.}{2011}]{2011MNRAS.418.1640W}
{Wu} J.,  {Zhou} X.,  {Ma} J.,   {Jiang} Z.,  2011, \mn@doi [\mnras]
  {10.1111/j.1365-2966.2011.19565.x}, \href
  {https://ui.adsabs.harvard.edu/abs/2011MNRAS.418.1640W} {418, 1640}

\bibitem[\protect\citeauthoryear{{Zhang}, {Zhou}, {Zhao}  \& {Dai}}{{Zhang}
  et~al.}{2015}]{2015RAA....15.1784Z}
{Zhang} B.-K.,  {Zhou} X.-S.,  {Zhao} X.-Y.,   {Dai} B.-Z.,  2015, \mn@doi
  [Research in Astronomy and Astrophysics] {10.1088/1674-4527/15/11/002}, \href
  {https://ui.adsabs.harvard.edu/abs/2015RAA....15.1784Z} {15, 1784}

\bibitem[\protect\citeauthoryear{{di Clemente}, {Giallongo}, {Natali},
  {Trevese}  \& {Vagnetti}}{{di Clemente} et~al.}{1996}]{1996ApJ...463..466D}
{di Clemente} A.,  {Giallongo} E.,  {Natali} G.,  {Trevese} D.,   {Vagnetti}
  F.,  1996, \mn@doi [\apj] {10.1086/177261}, \href
  {https://ui.adsabs.harvard.edu/abs/1996ApJ...463..466D} {463, 466}

\makeatother
\end{thebibliography}

\bsp	
\label{lastpage}
\end{document}